\documentclass[useAMS,usenatbib]{mn2e}
\usepackage{epsf,graphicx,rotating,url}

\title[On the dust geometry in radio-loud AGN]{On the dust geometry in radio-loud active galactic nuclei}

\author[H. Landt, C. L. Buchanan and P. Barmby]{Hermine Landt$^1$\thanks{E-mail: hlandt@unimelb.edu.au}, Catherine L. Buchanan$^{1}$ and
P. Barmby$^2$ \\ 
$^1$School of Physics, University of Melbourne, Parkville, VIC 3010, Australia \\ 
$^2$Department of Physics and Astronomy, University of Western Ontario, 1151 Richmond Street, London, ON N6A 3K7, Canada }

\begin{document}

\def\la{\mathrel{\hbox{\rlap{\hbox{\lower4pt\hbox{$\sim$}}}\hbox{$<$}}}}
\def\ga{\mathrel{\hbox{\rlap{\hbox{\lower4pt\hbox{$\sim$}}}\hbox{$>$}}}}

\date{Accepted ~~. Received ~~; in original form ~~}

\pagerange{\pageref{firstpage}--\pageref{lastpage}} \pubyear{2010}

\maketitle

\label{firstpage}

\begin{abstract}

  We use photometric and spectroscopic infrared observations obtained
  with the {\it Spitzer Space Telescope} of 12 radio-loud active
  galactic nuclei (AGN) to investigate the dust geometry. Our approach
  is to look at the change of the infrared spectral energy
  distribution (SED) and the strength of the 10 $\mu$m silicate
  feature with jet viewing angle. We find that (i) a combination of
  three or four blackbodies fits well the infrared SED; (ii) the
  sources viewed closer to the jet axis appear to have stronger warm
  ($\sim 300 - 800$ K) and cold ($\sim 150 - 250$ K) dust emissions
  relative to the hot component; and (iii) the silicate features are
  always in emission and strongly redshifted. We test clumpy torus
  models and find that (i) they approximate well the mid-infrared part
  of the SED, but significantly underpredict the fluxes at both near-
  and far-infrared wavelengths; (ii) they can constrain the dust
  composition (in our case to that of the standard interstellar
  medium); (iii) they require relatively large ($\sim 10\%-20\%$ the
  speed of light) redward displacements; and (iv) they give robust
  total mass estimates, but are insensitive to the assumed geometry.

\end{abstract}


\section{Introduction}

Some active galactic nuclei (AGN) exhibit in their optical spectra
both broad ($\sim 1\%-5\%$ the speed of light) and narrow emission
lines (type 1 AGN) and some only narrow (type 2 AGN). The existence of
apparently two distinct classes of AGN has been explained within
unified schemes by orientation effects: an optically thick, dusty
torus (or warped disc) located outside the accretion disc obscures the
broad emission line region for some lines of sight \citep[see reviews
by][]{Law87, Ant93, Urry95}. The strongest observational evidence
interpreted in favour of such a torus are broad emission lines in the
polarized, scattered light of numerous type 2 AGN
\citep[e.g.,][]{Ant85a, Coh99, Lum01} and significant infrared
emission in the continuua of most AGN. However, details of the
physical state of this torus, its precise geometry and location remain
unconstrained to this day.

The putative torus will absorb a significant fraction of the nuclear
(accretion) luminosity and reradiate it strongly in the infrared. The
dust radiative transfer problem was first investigated for a toroidal
geometry by \citet{Kro88} and \citet{Pier92, Pier93}. Because of the
difficulties in modelling a clumpy medium they assumed a uniform
density distribution, although they noted that the dust must be
concentrated in clouds to protect the grains. Their models reproduced
the gross features of the observed spectral energy distributions
(SEDs), but some major problems remained: (i) the torus was predicted
to emit anisotropically; (ii) the generated emission at far-infrared
wavelengths was insufficient; (iii) early observations of type 1 AGN
did not show the predicted strong 10 $\mu$m silicate emission features
\citep{Roche91, Clavel00, Spoon02}; and (iv) the required geometrical
thickness was difficult to support \citep{Kro07, Shi08}. These
problems persisted in all the following studies, which also employed
continuous density distributions \citep[e.g.,][]{Gra94, Efs95, Bem03}.

More recently, \citet{Nen02, Nen08a, Nen08b} have developed a
formalism for handling radiative transfer in clumpy media and have
applied it to AGN. Their model is able to reproduce some of the
observed SED features, if a small number of clouds along equatorial
rays is assumed. In particular, the infrared SEDs and strengths of the
silicate emission features are now predicted to be largely independent
of orientation and a large range of temperatures can coexist at a
given distance from the central source, as required by new
interferometric results \citep{Tri07, Tri09}. Nevertheless, the
situation is far from being resolved. Observations with the {\it
  Spitzer Space Telescope} have now started to show strong silicate
emission features in quasars and absorption troughs in sources viewed
edge-on, just as predicted by Pier \& Krolik \citep[e.g.,][]{Sieben05,
  Haas05, Hao05, Weed05, Buch06, Shi06, Ogle06, Cleary07, Hao07,
  Schweitzer08, Thomp09, Hiner09}.

Unravelling the structure of the dust obscuration is essential to our
understanding of AGN. In this paper we use for the first time both
photometric and spectroscopic {\it Spitzer Space Telescope}
observations of radio-loud AGN, for which actual viewing angles can be
determined, to investigate this topic. In Section 2 we introduce the
sample and discuss the method used to derive viewing angles. The
infrared data are presented in Section 3, whereas in Section 4 we
confront theory with observations. Finally, in Section 5 we conclude
and summarize our main results.

Throughout this paper we have assumed cosmological parameters $H_0=70$
km s$^{-1}$ Mpc$^{-1}$, $\Omega_{\rm M}=0.3$, and
$\Omega_{\Lambda}=0.7$.

\section{The sample} \label{sample}


\begin{table*}
\begin{turn}{90}
\begin{minipage}{240mm}
\vspace*{-0.5cm}
\caption{\label{general} General Properties of the Sample}
\begin{tabular}{lcccccccccccccccc}
\hline
Object Name & IAU Name & R.A.(J2000) & Decl.(J2000) & z & $\theta$ & $\mu_{\rm r}$ 
& $\sigma_{\mu_{\rm r}}$ & Ref. & $\beta_a$ & $\nu_{\rm m}$ & $F_{\rm m}$ 
& $\Theta_{\rm d}$ & Ref. & $F_{\rm x}$ & Ref. & $\delta$ \\
&&&&& [deg] & [mas/yr] & [mas/yr] &&& [GHz] & [Jy] & [mas] && [$\mu$Jy] && \\
(1) & (2) & (3) & (4) & (5) & (6) & (7) & (8) & (9) & (10) & (11) & (12) & (13) 
& (14) & (15) & (16) & (17) \\
\hline
III Zw 2$^{\star}$& 0007$+$106 & 00 10 31.0 & $+$10 58 30 & 0.090 & 28 & 0.166 & 0.016 & MOJAVE &  0.99 & 15 & 0.79 & 0.15 & B05 & 1.101 &Pic05& 1.9 \\
3C 47             & 0133$+$207 & 01 36 24.4 & $+$20 57 27 & 0.425 & 18 & 0.24  & 0.05  & VC94   &  6.21 &  8 & 0.07 & 0.20 & H02 & 0.363 & Ha06& 0.4 \\
3C 84             & 0316$+$413 & 03 19 48.1 & $+$41 30 42 & 0.018 & 31 & 0.266 & 0.050 & MOJAVE &  0.32 & 28 & 6.50 & 0.35 & R83 & 2.400 & E06 & 1.4 \\
S5 0615$+$820     & 0615$+$820 & 06 26 03.0 & $+$82 02 26 & 0.710 & 49 & 0.053 & 0.020 & MOJAVE &  2.12 &  2 & 0.86 & 3.50 & F00 & 0.040 & S97 & 0.4 \\
3C 207            & 0838$+$133 & 08 40 47.6 & $+$13 12 24 & 0.681 &  8 & 0.334 & 0.030 & MOJAVE & 12.94 &  8 & 0.60 & 0.28 & H02 & 0.120 & G03 & 3.0 \\
3C 245            & 1040$+$123 & 10 42 44.6 & $+$12 03 31 & 1.029 & 18 & 0.11  & 0.05  & VC94   &  5.87 & 11 & 0.59 & 0.33 & H87 & 0.160 & G03 & 1.7 \\
3C 263            & 1137$+$660 & 11 39 57.0 & $+$65 47 49 & 0.646 & 48 & 0.06  & 0.02  & VC94   &  2.23 &  8 & 0.12 & 0.37 & H02 & 0.290 & Ha02& 0.3 \\
3C 334            & 1618$+$177 & 16 20 21.8 & $+$17 36 24 & 0.555 & 34 & 0.10  & 0.03  & VC94   &  3.27 & 11 & 0.09 & 0.20 & H92 & 0.342 & Ha99& 0.4 \\
3C 336            & 1622$+$238 & 16 24 39.1 & $+$23 45 12 & 0.927 & 23 & 0.10  & 0.07  & H02    &  4.94 &  8 & 0.01 & 0.20 & H02 & 3.700 & P96 & 0.1 \\
4C $+$34.47       & 1721$+$343 & 17 23 20.8 & $+$34 17 58 & 0.206 & 30 & 0.28  & 0.03  & VC94   &  3.72 & 11 & 0.11 & 0.24 &Ho92a& 1.467 & Pa04& 0.2 \\
4C $+$28.45       & 1830$+$285 & 18 32 50.2 & $+$28 33 36 & 0.594 & 24 & 0.13  & 0.04  & VC94   &  4.50 &  5 & 0.50 & 0.30 &Ho92b& 0.309 & B97 & 0.8 \\
3C 390.3          & 1845$+$797 & 18 42 09.0 & $+$79 46 17 & 0.056 & 48 & 0.605 & 0.010 & MOJAVE &  2.26 & 11 & 0.41 & 0.59 & L81 & 7.600 & E06 & 0.1 \\
\hline
\end{tabular}

\parbox[]{23.5cm}{The columns are: (1) object name; (2) IAU name; (3)
  and (4) position, and (5) redshift from the NASA/IPAC Extragalactic
  Database (NED); (6) jet viewing angle calculated with the
  synchrotron self-Compton (SSC) formalism using the following
  observables: (7) proper motion and (8) $1\sigma$ error on proper
  motion, taken from reference in (9), where MOJAVE: large project for
  'Monitoring Of Jets in AGN with VLBA Experiments'
  \citep[e.g.,][]{Kel04, Lis09}, VC94: \citet{Verm94}, and H02:
  \citet{Hough02}; (10) apparent jet speed calculated from column (7);
  (11) synchrotron self-absorption frequency; (12) flux and (13) size
  at (11) of the core component; (14) reference for columns (11)-(13),
  where B05: \citet{Brun05}, F00: \citet{Fey00}, H87: \citet{Hough87},
  H92: \citet{Hough92}, H02: \citet{Hough02}, Ho92a: \citet{Hoo92a},
  Ho92b: \citet{Hoo92b}, L81: \citet{Lin81}, R83: \citet{Read83}; (15)
  X-ray flux at 1 keV, taken from the reference in (16), where B97:
  \citet{Bri97a}, E06: \citet{Evans06}, G03: \citet{Gam03}, Ha99:
  \citet{Har99}, Ha02: \citet{Har02}, Ha06: \citet{Har06}, P96:
  \citet{Prieto96}, Pa04: \citet{Page04}, Pic05: \citet{Pic05}, S97:
  \citet{Sam97}; and (17) relativistic Doppler factor calculated from
  columns (11)-(13) and (15).}

\medskip

\parbox[]{23.5cm}{$^{\star}$ position of the northern most source in a galaxy triplet}

\end{minipage}
\end{turn}
\end{table*}


The presence of strong radio jets in radio-loud AGN gives us a unique
opportunity to derive actual viewing angles, $\theta$ (defined as
the angle between the jet and the observer's line of sight), for this
object class. Following \citet{Ghi93}, we use here the method that
combines apparent jet speeds, $\beta_a$ (in units of the speed of
light), with relativistic Doppler factors, $\delta$, to calculate
viewing angles.

If $\beta_a$ is available from proper motion measurements and $\delta$
can be separately estimated, then the viewing angle is:

\begin{equation}
\label{angleeq}
\tan \theta = \frac{2 \beta_a}{{\beta_a}^2 + \delta^2 - 1}.
\end{equation}

\noindent
The largest uncertainty of this approach lies with the determination
of $\delta$, which involves several observables. Using the classical
condition that the (predicted) synchrotron self-Compton (SSC) flux
should not exceed the observed high-energy flux, $\delta$ can be
calculated as \citep{Marscher87}:

\begin{equation}
\label{deltaeq}
\delta = f(\alpha) F_{\rm m} \left[\frac{\ln(\nu_{\rm b}/\nu_{\rm m})}{F_{\rm x} \Theta_{\rm d}^{6+4\alpha}\nu_{\rm x}^{\alpha}\nu_{\rm m}^{5+3\alpha}}\right]^{1/(4+2\alpha)} (1+z),
\end{equation}

\noindent
where $\alpha$ is the spectral index of the thin synchrotron emission
(assumed to be 0.75), $\nu_{\rm b}$ is the synchrotron high-frequency
cut-off (assumed to be $10^{14}$ Hz), the function $f(\alpha) \simeq
0.08\alpha + 0.14$ (assumed to be 0.2), and the redshift $z$ is
known. The synchrotron flux $F_{\rm m}$ (in Jy) at the self-absorption
frequency $\nu_{\rm m}$ (in Hz) of the core component with angular
size $\Theta_{\rm d}$ (in mas) need to be determined from appropriate
Very Large Baseline Interferometry (VLBI) observations. Observables
are also the high-frequency (X-ray) flux $F_{\rm x}$ (in Jy) at the
(X-ray) frequency $\nu_{\rm x}$ (in keV) of this component.

The most critical parameters in eq. (\ref{deltaeq}) are the VLBI
observables $\nu_{\rm m}$ and $\Theta_{\rm d}$, and these are in
general difficult and expensive to determine. The determination of
$\nu_{\rm m}$ ideally requires a VLBI spectrum (i.e., high spatial
resolution observations at different radio frequencies) of the
strongest (core) component or a sophisticated spectral decomposition
of a low-resolution spectrum to isolate the different (VLBI)
components. On the other hand, observationally $\Theta_{\rm d}$ will
depend on the frequency itself (the higher the frequency, the higher
the spatial resolution) and, therefore, will be an upper limit
only. This then makes $\delta$ strictly speaking a lower limit and so
the viewing angle an upper limit.

As \citet{Ghi93} \citep[and also][]{Rok03} have shown, this method
samples a wide range of viewing angles, including the regime of
$20^\circ<\theta<60^\circ$ that we are interested in. \citet{Verm94}
presented a large compilation of sources with measured jet proper
motions. We have updated their list with later literature and have
calculated viewing angles for all the sources. From the updated list,
and restricting ourselves to $z\la1$ in order to keep the emitted
far-infrared accessible, we have chosen for observations with the {\it
  Spitzer Space Telescope} all sources with viewing angles $\theta \ga
20^\circ$ (12 objects; see Table \ref{general}). We note that the
source 3C 207 no longer obeys this criterion based on most recent
proper motion data, but is kept in the sample for comparison. The cut
in viewing angle avoids sources with infrared emission dominated by
the relativistically beamed jet, so allowing for a reliable
determination of the dust SED. All selected sources are radio quasars
(i.e., type 1 AGN), except for 3C 84 (= NGC 1275), which is a radio
galaxy (i.e., type 2 AGN).

The total errors in our viewing angles are difficult to determine,
since the errors in the Doppler factors $\delta$ are largely
unconstrained. But, given the errors in the proper motions (Table
\ref{general}, column (8)), we derive an average lower limit on them
of $\sim 50\%$. In this respect, we note that \citet{Rok03} argue that
it is unlikely that Doppler factors estimated using
eq. (\ref{deltaeq}) are wrong by a very large factor, since they
observe excellent correlations between $\delta$ and several emission
line properties. We concur with this conclusion given our results in
Section \ref{blackbody}. In addition, a consistency check that we
carry out in Section \ref{jet} supports the notion that the trend in
viewing angle that we obtain for our sample is qualitatively correct.

\section{Observations and data reduction}

We observed our sample with the {\it Spitzer Space Telescope} in Cycle
1 (ID: 3551). We used all its instruments and imaged the sources with
the Infrared Array Camera \citep[IRAC;][]{irac} in four bands (3.6,
4.5, 5.8, and 8.0 $\mu$m) and with the Multiband Imaging Photometer
for Spitzer \citep[MIPS;][]{mips} in three bands (24, 70, and 160
$\mu$m). In addition we obtained low-resolution spectroscopy with the
Infrared Spectrograph \citep[IRS;][]{irs} in staring mode using the
appropriate short and long order modules in order to cover well the
rest-frame wavelength region around \mbox{10 $\mu$m}.

In Tables \ref{spitzerphot} and \ref{spitzerspec} we list the details
for the photometric and spectroscopic observations, respectively. Some
of the approved observations were allocated to other programs and were
accessible to us only after the proprietary period expired. We also
note that we now include in this study all low-resolution (staring
mode) IRS spectra available in the archive for our sources and not
only those initially requested. In the following we describe the data
reduction process and measurements.

\subsection{The photometry}

\begin{table*}
\caption{\label{spitzerphot} 
{\sl Spitzer} Photometry Journal of Observations}
\begin{tabular}{lrlcrlcrlcrlc}
\hline
Object Name & \multicolumn{3}{c}{IRAC} & \multicolumn{9}{c}{MIPS} \\
& ID & observation & texp. & \multicolumn{3}{c}{24 $\mu$m} & \multicolumn{3}{c}{70 $\mu$m} & \multicolumn{3}{c}{160 $\mu$m} \\ 
&& date & [sec] & ID & observation & texp. & ID & observation & texp. & ID & observation & texp.\\
&&&&& date & [sec] && date & [sec] && date & [sec] \\
(1) & (2) & (3) & (4) & (5) & (6) & (7) & (8) & (9) & (10) & (11) & (12) & (13) \\
\hline
III Zw 2      & 3551 & 2004 Dec 15 & 4$\times$~2 &   86 & 2004 Dec 26 & 2$\times$~3 &   86 & 2004 Dec 26 & 2$\times$~3 & 3551 & 2005 Aug 2  & 10$\times$10 \\
3C 47         & 3551 & 2005 Jan 16 & 4$\times$~2 & 3551 & 2005 Jan 29 & 3$\times$~3 & 3551 & 2005 Jan 29 & 3$\times$10 & 3551 & 2005 Jan 29 & 10$\times$10 \\
3C 84         & 3228 & 2005 Feb 20 & 5$\times$30 & 3551 & 2005 Feb 26 & 3$\times$~3 & 3551 & 2005 Feb 26 & 3$\times$~3 & 3418 & 2005 Feb 1  & ~5$\times$10 \\
S5 0615$+$820 & 3551 & 2004 Nov 20 & 4$\times$12 & 3551 & 2005 Mar 3  & 3$\times$~3 & 3551 & 2005 Mar 3  & 3$\times$10 & 3551 & 2005 Mar 3  & 10$\times$10 \\
3C 207        & 3551 & 2005 May 10 & 4$\times$12 &   74 & 2005 Apr 10 & 1$\times$10 &   74 & 2005 Apr 10 & 4$\times$10 &   74 & 2005 Apr 10 & ~4$\times$10 \\
3C 245        &40072 & 2008 Jun 10 & 1$\times$30 &40072 & 2008 Jan 6  & 1$\times$10 &&&&&& \\
3C 263        & 3551 & 2004 Nov 1  & 4$\times$~2 &   74 & 2005 Apr 9  & 1$\times$~3 &   74 & 2005 Apr 9  & 1$\times$10 &   74 & 2005 Apr 9  & ~4$\times$~3 \\
3C 334        & 3551 & 2005 Mar 26 & 4$\times$~2 &   74 & 2005 Apr 7  & 1$\times$10 &   74 & 2005 Apr 7  & 1$\times$10 &   74 & 2005 Apr 7  & ~4$\times$10 \\
3C 336        & 3551 & 2005 Mar 27 & 4$\times$12 &   74 & 2005 Apr 12 & 1$\times$10 &   74 & 2005 Apr 12 & 6$\times$10 &   74 & 2005 Apr 12 & ~4$\times$10 \\
4C $+$34.47   & 3551 & 2005 Mar 30 & 4$\times$~2 & 3551 & 2005 Apr 6  & 3$\times$~3 & 3551 & 2005 Apr 6  & 3$\times$10 & 3551 & 2005 Apr 6  & 10$\times$10 \\
4C $+$28.45   & 3551 & 2004 Oct 8  & 4$\times$~2 & 3551 & 2004 Oct 17 & 3$\times$~3 & 3551 & 2004 Oct 17 & 3$\times$10 & 3551 & 2004 Oct 17 & 10$\times$10 \\
3C 390.3      &50763 & 2008 May 13 & 9$\times$12 & 3327 & 2004 Oct 17 & 4$\times$10 & 3551 & 2004 Oct 18 & 3$\times$10 & 3551 & 2004 Oct 18 & 10$\times$10 \\
\hline
\end{tabular}

\medskip

\parbox[]{19cm}{The columns are: (1) object name; for photometry with
  the Infrared Array Camera (IRAC) (2) program number, (3) observation
  date and (4) exposure time; for photometry with the Multiband
  Imaging Photometer for Spitzer (MIPS) (5) program number, (6)
  observation date, and (7) exposure time in the 24 $\mu$m band, (8)
  program number, (9) observation date, and (10) exposure time in the
  70 $\mu$m band, and (11) program number, (12) observation date, and (13)
  exposure time in the 160 $\mu$m band.}

\end{table*}

The IRAC data were processed with the {\it Spitzer} Science Center
(SSC) pipeline v14.0, except for 3C 84 and 3C 390.3, for which we used
the pipeline v18.7.0 since it corrects artifacts near bright
sources. We measured flux densities from 3.6 to 8.0 $\mu$m on the
`post-basic calibrated data (BCD)' mosaics produced by the
pipeline. For 3C 84 and 3C 390.3, mosaics made from the short-exposure
images taken in high-dynamic range mode were used. We used aperture
photometry and adopted a 10-pixel (12 arcsec) source radius, with
background counts estimated in a 10-20 pixel (12-24 arcsec) radius
annulus and subtracted. The sources 3C 336 and S5 0615$+$820 had other
objects nearby, therefore, we measured their fluxes in smaller 5-pixel
(6 arcsec) apertures.

We note that the 3.6 and 4.5 $\mu$m images of III~Zw~2 show that this
object may be a blend of two very close sources, and so flux densities
at these wavelengths could be overestimated. The 3.6 and 4.5 $\mu$m
images of 3C 84 probably are contaminated by an extended component
since no separate central point source is visible. For the source 3C
390.3 two data sets were available separated by about four years and
we have considered both. However, the observations contemporaneous
with the MIPS photometry and IRS spectroscopy were severely affected
by saturation, and, therefore, not useful for our purpose.

All IRAC flux densities were aperture-corrected using the values in
the IRAC Data Handbook (v3.0, 2006). No `array-location-dependent'
photometric correction was applied, since these are red sources and do
not require such a correction. Table \ref{phot} lists our results. All
sources have been detected in all four IRAC bands. We give $1\sigma$
uncertainties estimated using the standard IRAF photometry formula. We
note that these values do not include the error in absolute
calibration \citep[a few per cent;][]{iracerr}.

The MIPS data were reprocessed with the SSC pipeline v16.1.0.
Measurements on 24 $\mu$m images followed standard procedures using
the post-BCD pipeline mosaics. We measured flux densities in the
standard aperture ($35''$ radius) with the recommended aperture
correction of 1.082 applied, except in the case of 3C 336 where a
$7''$ radius aperture with a correction factor of 1.61 were used. The
flux density measured for 3C 84 (2.9 Jy) is formally above the
saturation limit for 3-sec. exposures, but the object is only
`soft-saturated' and the pipeline correctly replaced the saturated
pixels with values from the 0.5-sec. exposures taken as part of the
photometry AOR. Table \ref{phot} lists the results. All sources have
been detected. We have computed uncertainties using the same method as
for IRAC, which again do not include the errors in the absolute
calibration \citep[$\sim 2\%$;][]{mips24err}.

For the measurements at 70 $\mu$m we used mostly the pipeline-produced
`filtered' mosaics, although in a few cases we remade the mosaics
using time- and column-filtering of the BCDs to remove negative
sidelobes around bright sources. Per the MIPS Data Handbook (v3.3.1)
we measured the flux densities in $35''$ radius apertures with sky
annuli of 39-60 arcsec and an aperture correction of 1.22. Our results
are listed in Table \ref{phot}. Following \citet{Carp08}, we
calculated uncertainties in the flux density using a similar equation
to that for IRAC and MIPS-24, but neglecting Poisson noise and
applying instead multiplicative factors to account for the noise
correlation between pixels due to resampling during mosaicing and
excess sky noise due to the data-taking procedure. The absolute
calibration uncertainty at 70 $\mu$m is $5\%$ \citep{mips70err}, again
not included in the tabulated values. The source 3C 84 is very bright
at 70 $\mu$m but, as at 24 $\mu$m, is just below the saturation
limit. Five sources were not detected at 70 $\mu$m and we give only
$3\sigma$ upper limits. One source (3C 245) did not have MIPS-70 data.

For the measurements at 160 $\mu$m we used the pipeline-produced
`filtered' BCDs and made mosaics using the MOPEX software
\citep{mopex}. However, for the bright source 3C 84 and the source 3C
390.3 that has another bright object nearby we produced mosaics from
the unfiltered images. Following the MIPS Data Handbook, we measured
flux densities in $48''$ apertures with sky annuli of 64-128 arcsec
and an aperture correction of 1.60. The only exception was the source
3C 390.3. Due to the presence of a nearby brighter source separated by
$\sim 1$ arcmin, we measured its flux density instead in an aperture
of $32"$ with an aperture correction of 1.97. Our results are listed
in Table \ref{phot}. Uncertainties for all sources were calculated as
described above for the MIPS-70 data. The absolute calibration
uncertainty at 160 $\mu$m is $12\%$ \citep{mips160err} and is not
included in the tabulated values. The source 3C 84 is very bright at
160 $\mu$m and at the saturation limit, therefore, its flux density is
highly uncertain and is taken to be $20\%$ \citep{mips160err}. Seven
sources were not detected at 160 $\mu$m and we give only $3\sigma$
upper limits. One source (3C 245) did not have MIPS-160 data.

The archival MIPS data we used for four objects, namely, 3C 207, 3C
263, 3C 334 and 3C 336, were analysed also by \citet{Cleary07}. Our
flux densities in the 24 $\mu$m , 70 $\mu$m and 160 $\mu$m bands are
consistent with theirs within $2 \sigma$, however, contrary to these
authors we regard the sources 3C 207 and 3C 263 as undetected in the
MIPS-70 images and the sources 3C~334 and 3C~336 as detected in the
MIPS-160 images.

\subsection{The spectroscopy} \label{spec}

\begin{table*}
\caption{\label{spitzerspec} 
{\sl Spitzer} Spectroscopy Journal of Observations}
\begin{tabular}{lrlcccrlccc}
\hline
Object Name & ID & observation & peak-up & SL2   & SL1   & ID & observation & peak-up & LL2 & LL1 \\ 
            &    & date        &         & texp. & texp. &    & date        &         & texp. & texp. \\
            &    &             &         & [sec] & [sec] &    &             &         & [sec] & [sec] \\
(1) & (2) & (3) & (4) & (5) & (6) & (7) & (8) & (9) & (10) & (11) \\
\hline
III Zw 2      &   86 & 2005 Jul 10 & blue & 4$\times$~30 & 2$\times$~60 &&&&& \\
3C 47         & 3551 & 2005 Jan 11 & none & --           & 4$\times$~14 & 3551 & 2005 Jan 11 & none & 4$\times$120 & -- \\
3C 84         &   14 & 2004 Aug 30 & blue & 2$\times$~14 & 2$\times$~14 &   14 & 2004 Aug 30 & blue & 4$\times$~~6 & ~4$\times$~~6 \\
S5 0615$+$820 & 3551 & 2004 Oct 23 & none & --           & 4$\times$~60 & 3551 & 2004 Oct 23 & none & 4$\times$120 & -- \\
3C 207        & 3551 & 2004 Apr 15 & none & --           & 4$\times$~60 &   74 & 2005 Nov 15 & blue & 4$\times$120 & ~4$\times$120 \\
3C 245        &      &             &      &              &              &40314 & 2008 Jan 10 & none & --           & 12$\times$120 \\
3C 263        & 3551 & 2004 Nov 16 & none & --           & 4$\times$~60 &   74 & 2004 Apr 14 & blue & 4$\times$120 & ~4$\times$120 \\
3C 334        & 3551 & 2005 Mar 14 & none & --           & 4$\times$~14 &   74 & 2004 Jul 16 & blue & 4$\times$120 & ~4$\times$120 \\
3C 336        &      &             &      &              &              &   74 & 2004 Mar 22 & blue & 4$\times$120 & ~4$\times$120 \\
4C $+$34.47   & 3551 & 2005 Mar 14 & none & --           & 4$\times$~~6 & 3551 & 2005 Mar 14 & none & 4$\times$~14 & -- \\
4C $+$28.45   & 3551 & 2004 Oct 23 & none & --           & 4$\times$~60 & 3551 & 2004 Oct 23 & none & 4$\times$120 & -- \\
3C 390.3      &   82 & 2004 Mar 24 & blue & 1$\times$240 & 1$\times$240 &   82 & 2004 Mar 24 & blue & 4$\times$120 & ~4$\times$120 \\
\hline
\end{tabular}

\medskip

\parbox[]{16.5cm}{The columns are: (1) object name; for low-resolution
  spectroscopy with the Infrared Spectrograph (IRS) short order
  modules (2) program number, (3) observation date, (4) peak-up
  imaging array and exposure times for the slits (5) SL2 (5.2-8.7
  $\mu$m) and (6) SL1 (7.4-14.5 $\mu$m); for low-resolution
  spectroscopy with the IRS long order modules (7) program number, (8)
  observation date, (9) peak-up imaging array and exposure times for
  the slits (10) LL2 (14.0-21.3 $\mu$m), and (11) LL1 (19.5-38.0
  $\mu$m).}

\end{table*}

The raw data were processed through the SSC pipeline v15.3 for the
short-low (SL) modules and v17.2 for the long-low (LL) modules. The
BCDs produced by the pipeline were cleaned for rogue pixels using the
IRSCLEAN software\footnote{IRSCLEAN was written by the IRS GTO team
  (G. Sloan, D. Devost, \& B. Sargent). It is distributed by the {\it
    Spitzer} Science Center at Caltech.}, then spectra were extracted
using the SMART software \citep{smart}. Where multiple exposures were
obtained, the two-dimensional spectral images were median-combined.

Two positions (nods) were observed for each target, and sky
subtraction was performed on the images, by subtracting the off-source
nod position image from the on-source image for each module. Spectra
were extracted and the fluxes calibrated using the default tapered
column apertures for each module. The modules at each nod position
were merged and the edges and overlapping regions trimmed. The spectra
from the two nod positions were then averaged to produce the final
spectrum. Uncertainty images provided by the pipeline and propagated
through SMART were used to produce the uncertainties on the final
spectra. Our results are shown in Fig. \ref{sed}.

\begin{table*}
\caption{\label{phot} 
{\sl Spitzer} Photometry Results}
\begin{tabular}{lcccccccccccccc}
\hline
Object Name & F$_{3.6}$ & $\sigma_{3.6}$ & F$_{4.5}$ & $\sigma_{4.5}$ & F$_{5.8}$ & $\sigma_{5.8}$ & F$_{8.0}$ & $\sigma_{8.0}$ &  F$_{24}$ & $\sigma_{24}$ &  F$_{70}$ & $\sigma_{70}$ & F$_{160}$ & $\sigma_{160}$ \\
& [mJy] & [mJy] & [mJy] & [mJy] & [mJy] & [mJy] & [mJy] & [mJy] & [mJy] & [mJy] & [mJy] & [mJy] & [mJy] & [mJy] \\
\hline
III Zw 2      & 22.2  & 0.1  & 28.1  & 0.1  &  35.3  & 0.4  &  48.8  & 0.2  &  138   & 1   &  128 &  17 &   59 &   6 \\
3C 47         &  5.3  & 0.1  &  6.9  & 0.1  &   9.2  & 0.3  &  11.2  & 0.2  &   36.8 & 0.4 &   37 &  10 &$<$27 & \\
3C 84         & 75.81 & 0.09 & 89.12 & 0.06 & 132.2  & 0.2  & 291.1  & 0.2  & 2930   & 3   & 3990 &  33 & 4350 & 870 \\
S5 0615$+$820 &  0.63 & 0.01 &  0.78 & 0.01 &   1.00 & 0.05 &   1.24 & 0.02 &    4.8 & 0.3 &$<$18 &     &$<$21 & \\
3C 207        &  1.54 & 0.01 &  2.09 & 0.02 &   2.75 & 0.07 &   3.94 & 0.07 &   13.2 & 0.4 &$<$26 &     &$<$36 & \\
3C 245        &  1.54 & 0.01 &  2.12 & 0.01 &   3.66 & 0.03 &   5.55 & 0.03 &   20.9 & 0.6 &&&&\\
3C 263        &  5.5  & 0.1  &  7.3  & 0.1  &   8.9  & 0.3  &  11.5  & 0.2  &   26.7 & 0.7 &$<$35 &     &$<$18 & \\
3C 334        &  3.44 & 0.09 &  4.7  & 0.1  &   5.5  & 0.3  &   7.4  & 0.2  &   36.0 & 0.3 &   83 &  11 &   28 &   7 \\
3C 336        &  0.76 & 0.01 &  1.04 & 0.01 &   1.44 & 0.03 &   1.93 & 0.03 &    3.9 & 0.1 &   19 &   5 &   62 &  10 \\
4C $+$34.47   &  9.0  & 0.1  & 11.9  & 0.1  &  13.8  & 0.3  &  18.2  & 0.2  &   59.9 & 0.5 &$<$21 &     &$<$15 & \\
4C $+$28.45   &  2.19 & 0.09 &  3.4  & 0.1  &   3.8  & 0.3  &   5.4  & 0.2  &   18.2 & 0.4 &$<$24 &     &$<$33 & \\
3C 390.3      & 56.0  & 0.1  & 66.2  & 0.1  &  73.1  & 0.2  &  89.0  & 0.2  &  234.0 & 0.2 &  120 &   6 &$<$24 & \\
\hline
\end{tabular}
\end{table*}

\section{The dust geometry}


\begin{table*}
\caption{\label{suppl} 
Two Micron All-Sky Survey (2MASS) Fluxes}
\begin{tabular}{lcccccc}
\hline
Object Name & $J$ (1.235 $\mu$m) & $\sigma_J$ & $H$ (1.622 $\mu$m) & $\sigma_H$ & $K_s$ (2.159 $\mu$m) & $\sigma_{K_s}$ \\ 
& [mJy] & [mJy] & [mJy] & [mJy] & [mJy] & [mJy] \\
\hline
III Zw 2      &  4.5  & 0.2  &  6.3  & 0.3  & 12.9  & 0.3  \\
3C 47         &  0.46 & 0.05 &  0.68 & 0.07 &  1.01 & 0.09 \\
3C 84         & 12.4  & 0.9  & 16.6  & 1.2  & 20    & 1    \\
S5 0615$+$820 &$<$0.13&      &$<$0.22&      &$<$0.26&      \\ 
3C 207        &  0.34 & 0.05 &  0.42 & 0.06 &  0.64 & 0.07 \\
3C 245        &  0.40 & 0.05 &  0.41 & 0.08 &  0.51 & 0.08 \\
3C 263        &  1.87 & 0.07 &  1.79 & 0.08 &  2.30 & 0.09 \\
3C 334        &  0.96 & 0.05 &  1.10 & 0.08 &  1.54 & 0.08 \\
3C 336        &  0.40 & 0.04 &  0.33 & 0.07 &  0.45 & 0.08 \\
4C $+$34.47   &  2.72 & 0.06 &  3.03 & 0.09 &  4.7  & 0.1  \\
4C $+$28.45   &  0.32 & 0.05 &  0.53 & 0.07 &  0.73 & 0.06 \\
3C 390.3      &  5.8  & 0.2  &  9.4  & 0.3  & 13.7  & 0.4  \\
\hline
\end{tabular}
\end{table*}

\begin{figure*}
\centerline{
\includegraphics[scale=0.28]{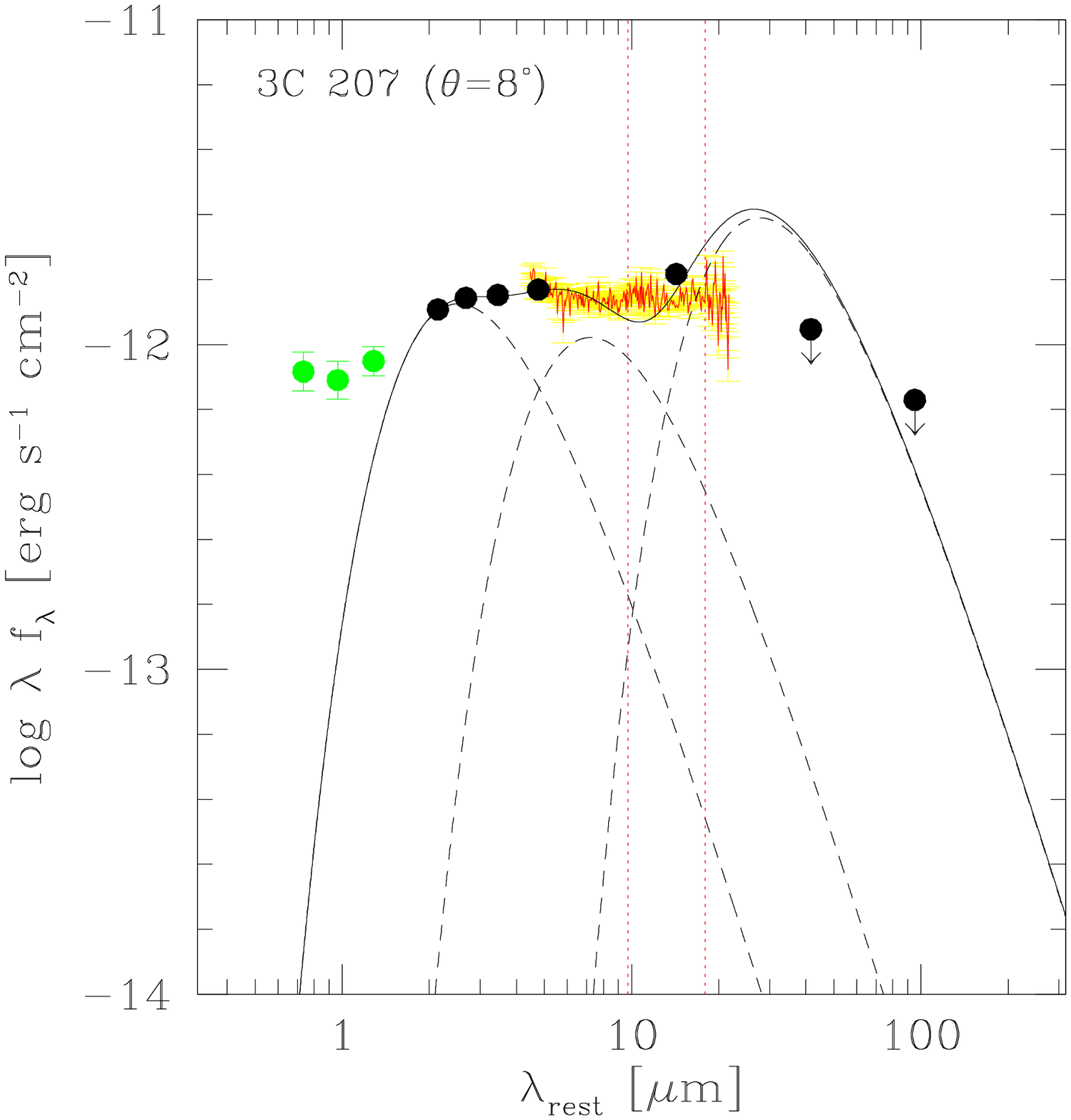} 
\includegraphics[scale=0.28]{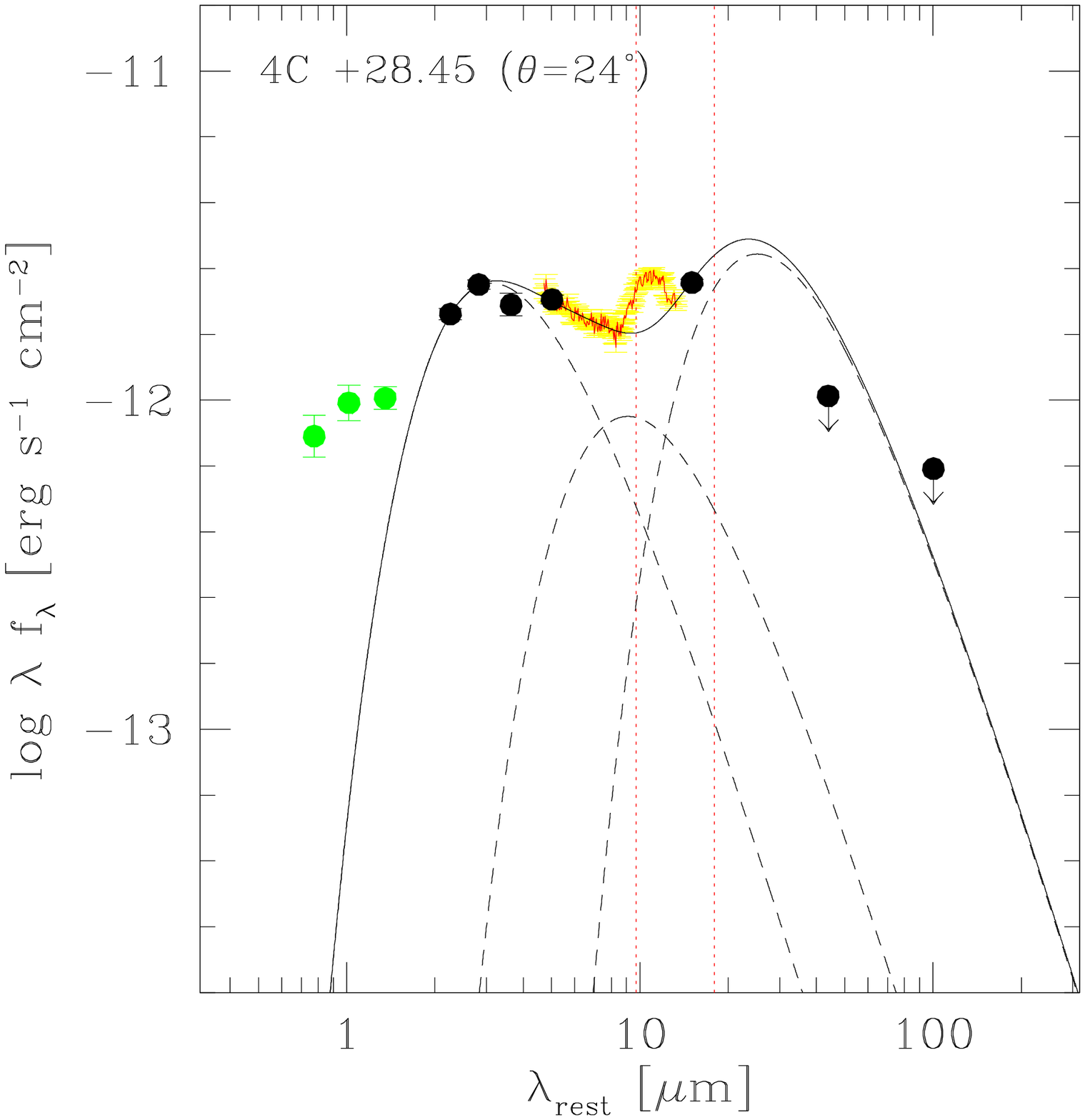}
\includegraphics[scale=0.28]{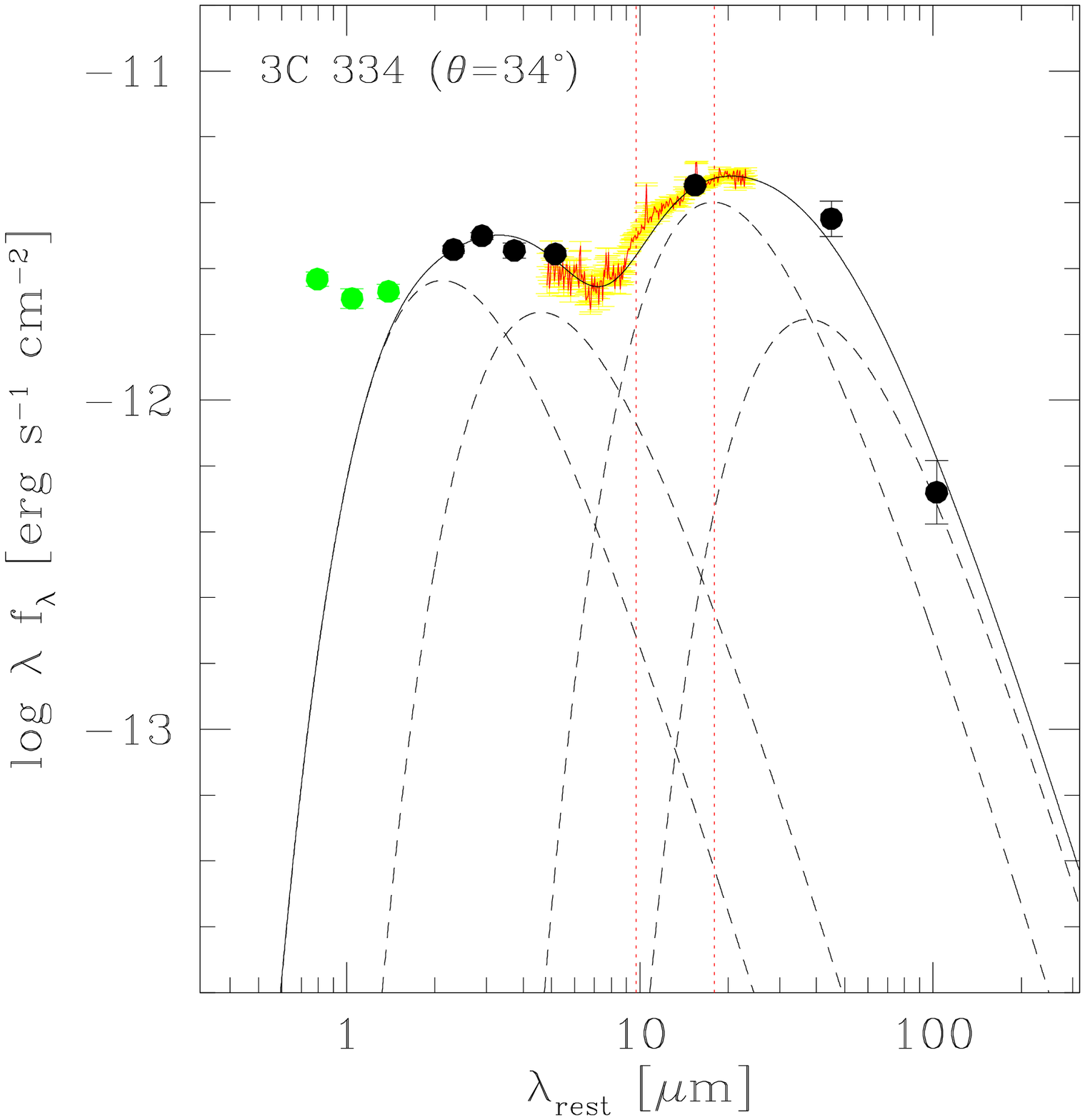}
}
\centerline{
\includegraphics[scale=0.28]{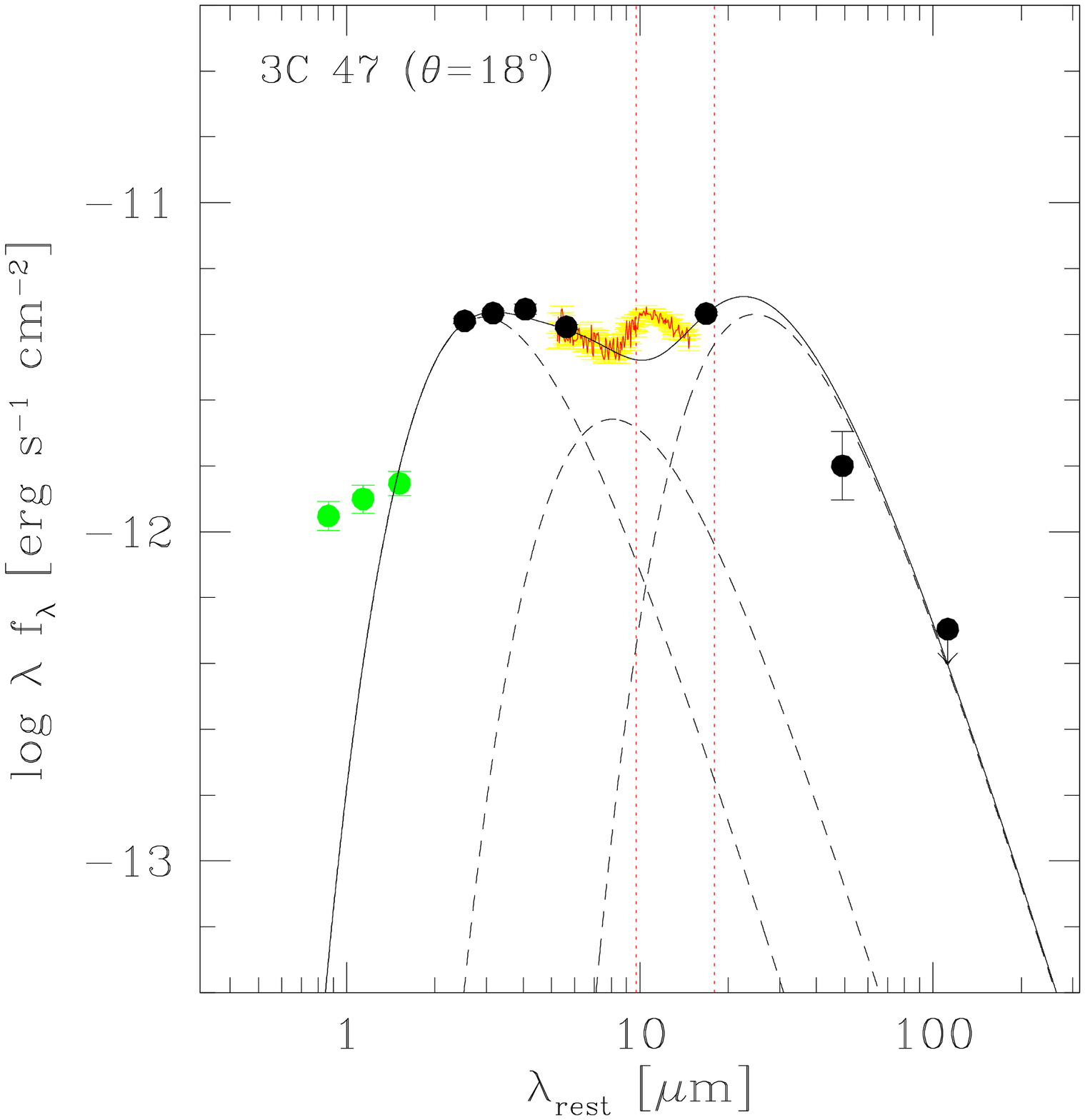}
\includegraphics[scale=0.28]{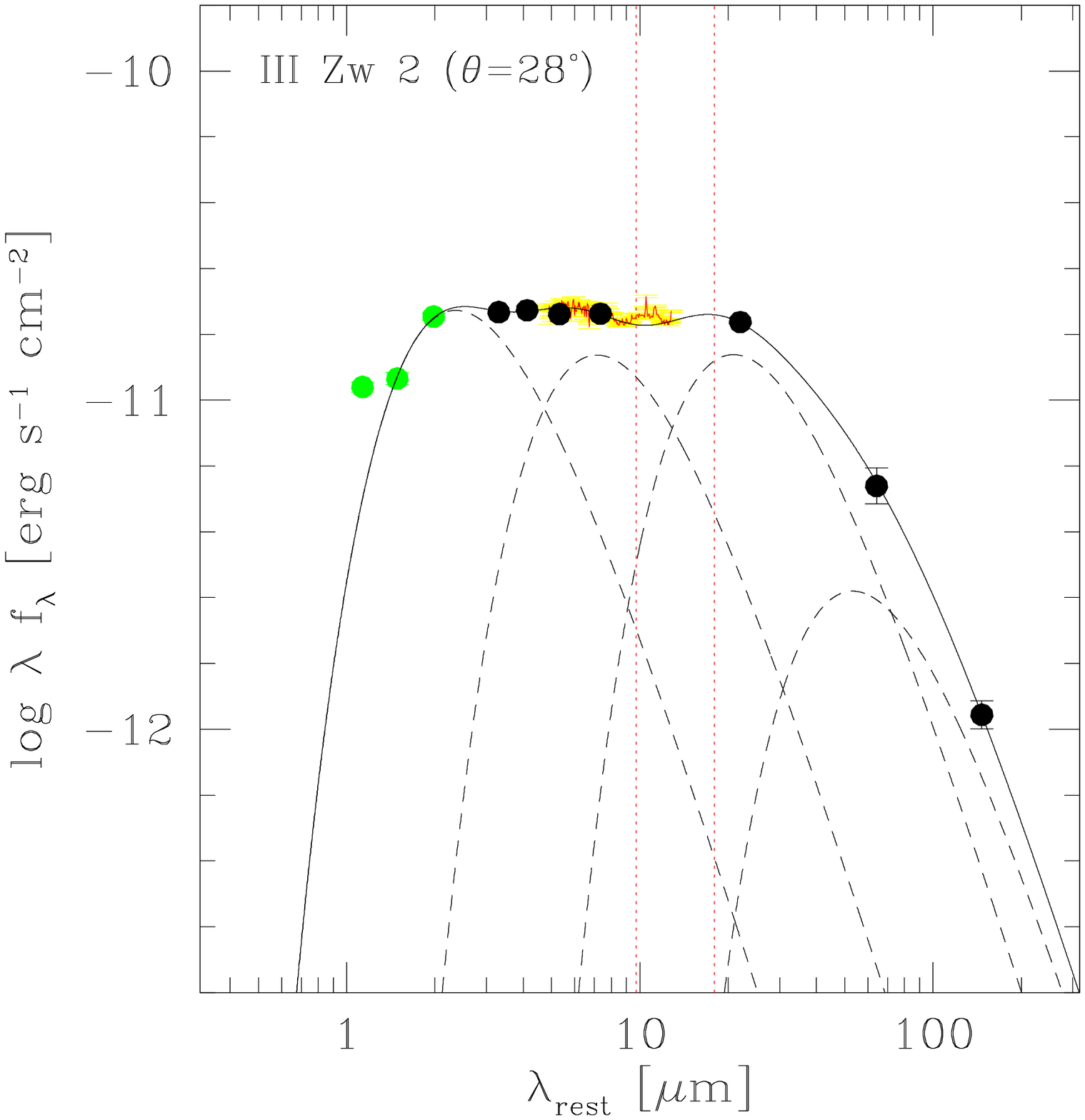}
\includegraphics[scale=0.28]{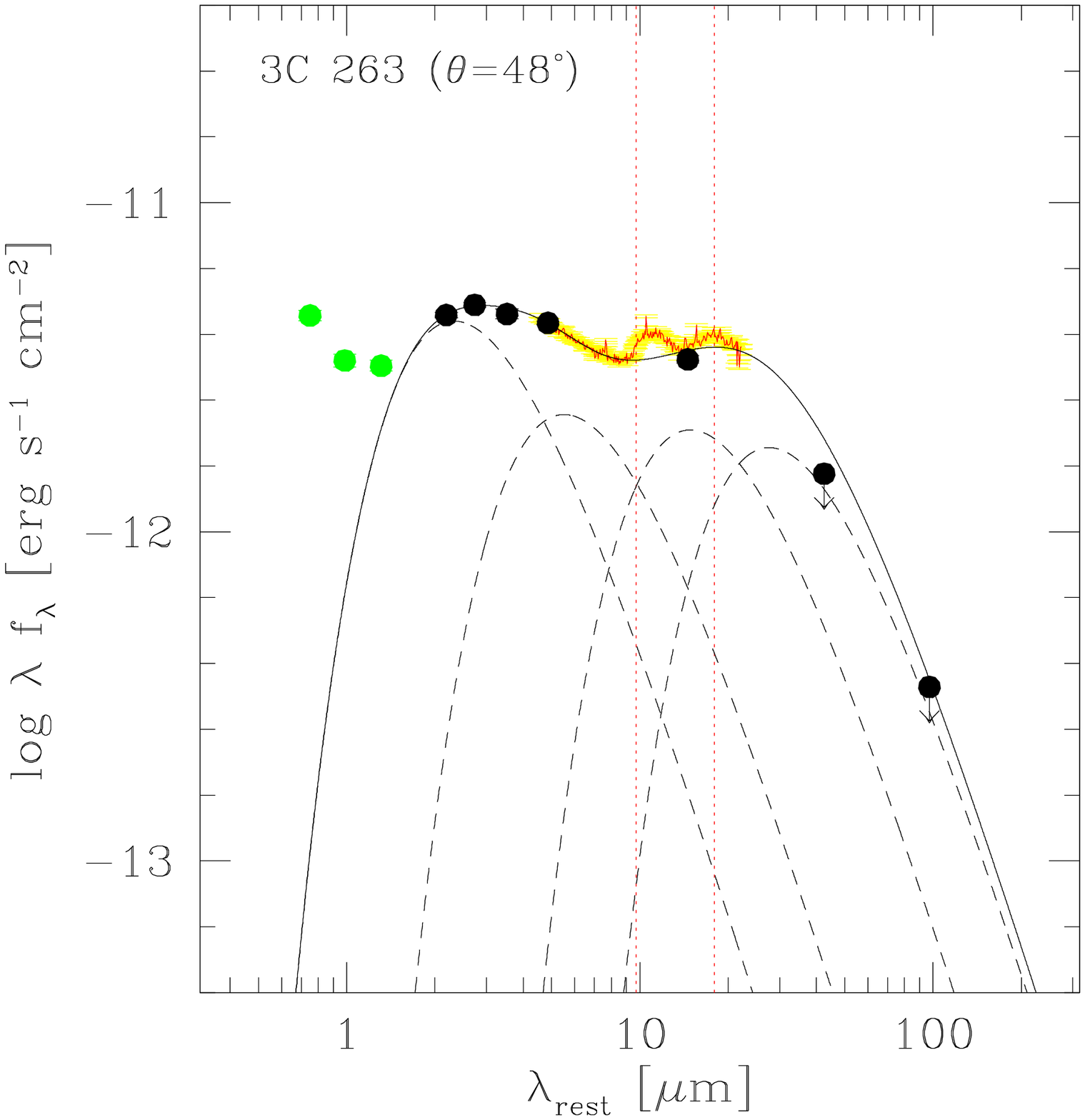}
}
\centerline{
\includegraphics[scale=0.28]{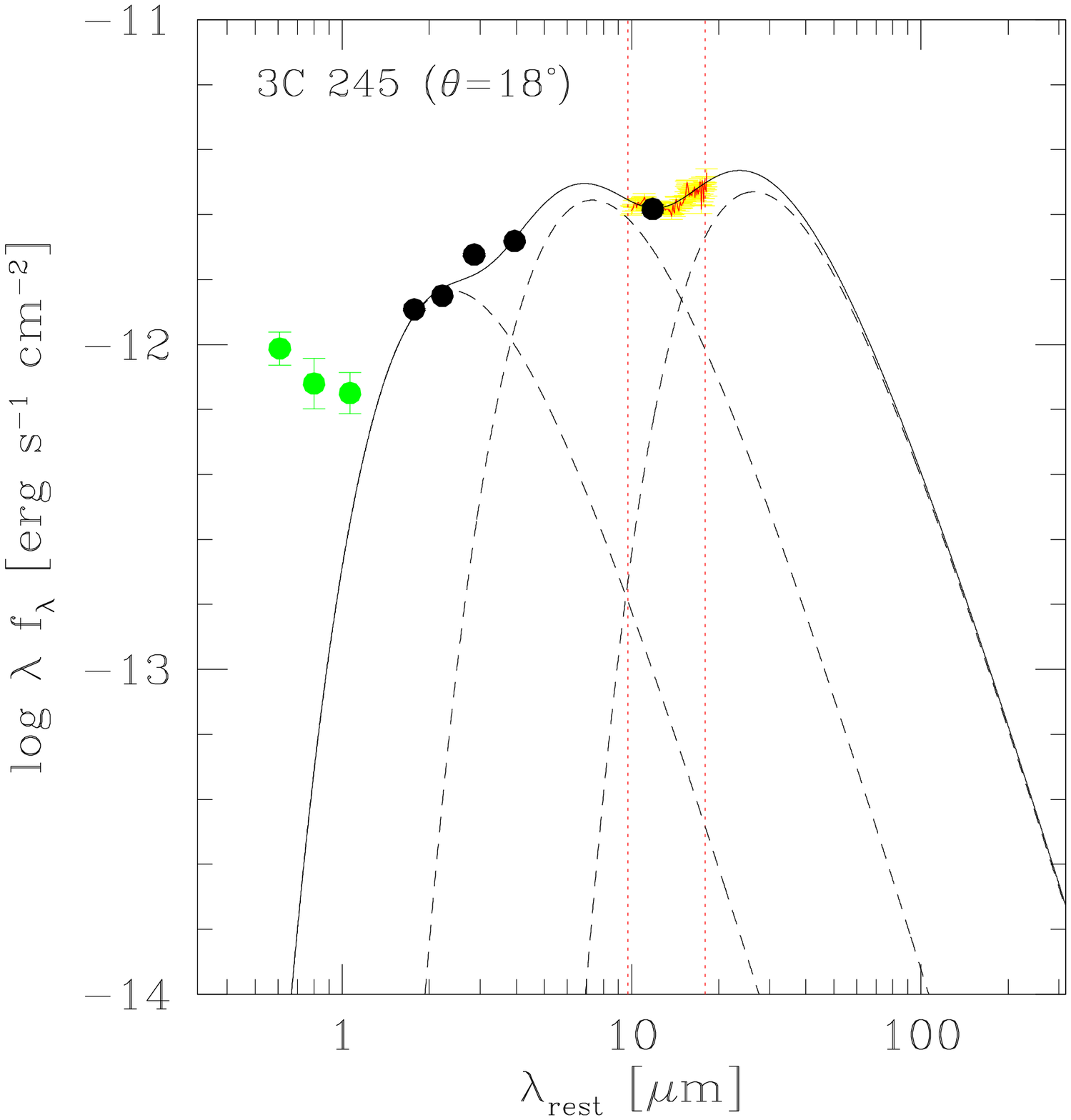}
\includegraphics[scale=0.28]{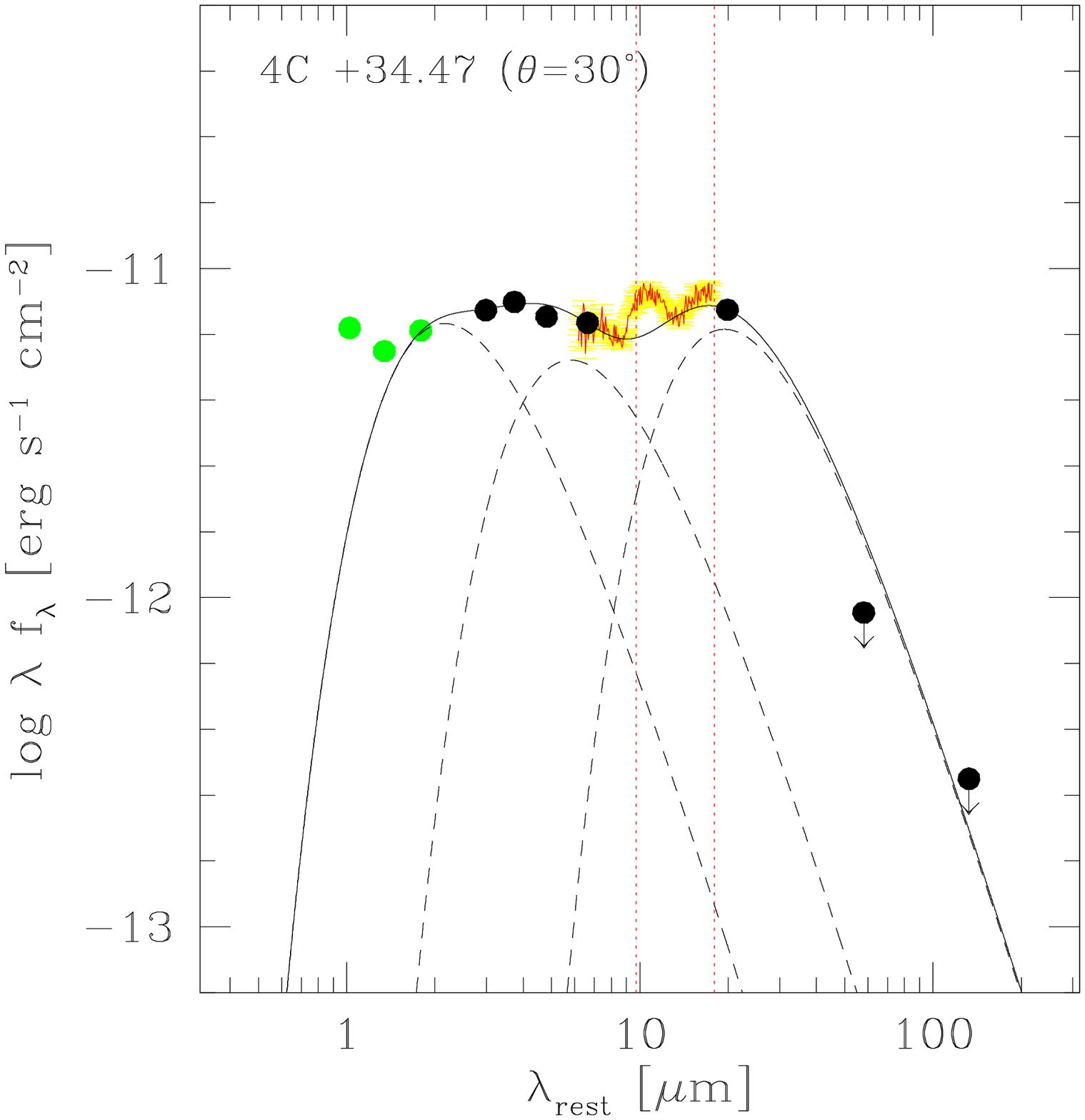}
\includegraphics[scale=0.28]{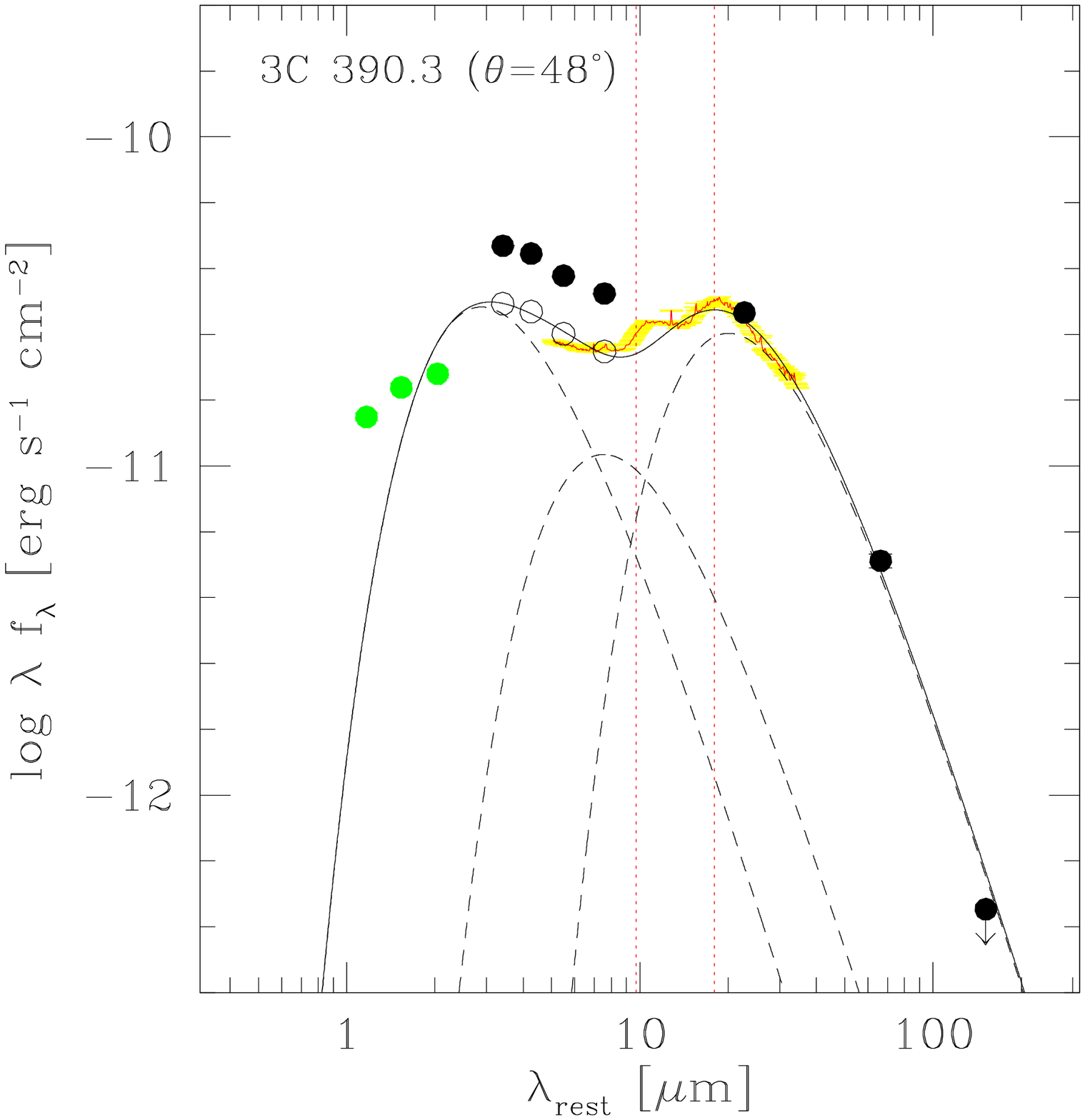}
}
\centerline{
\includegraphics[scale=0.28]{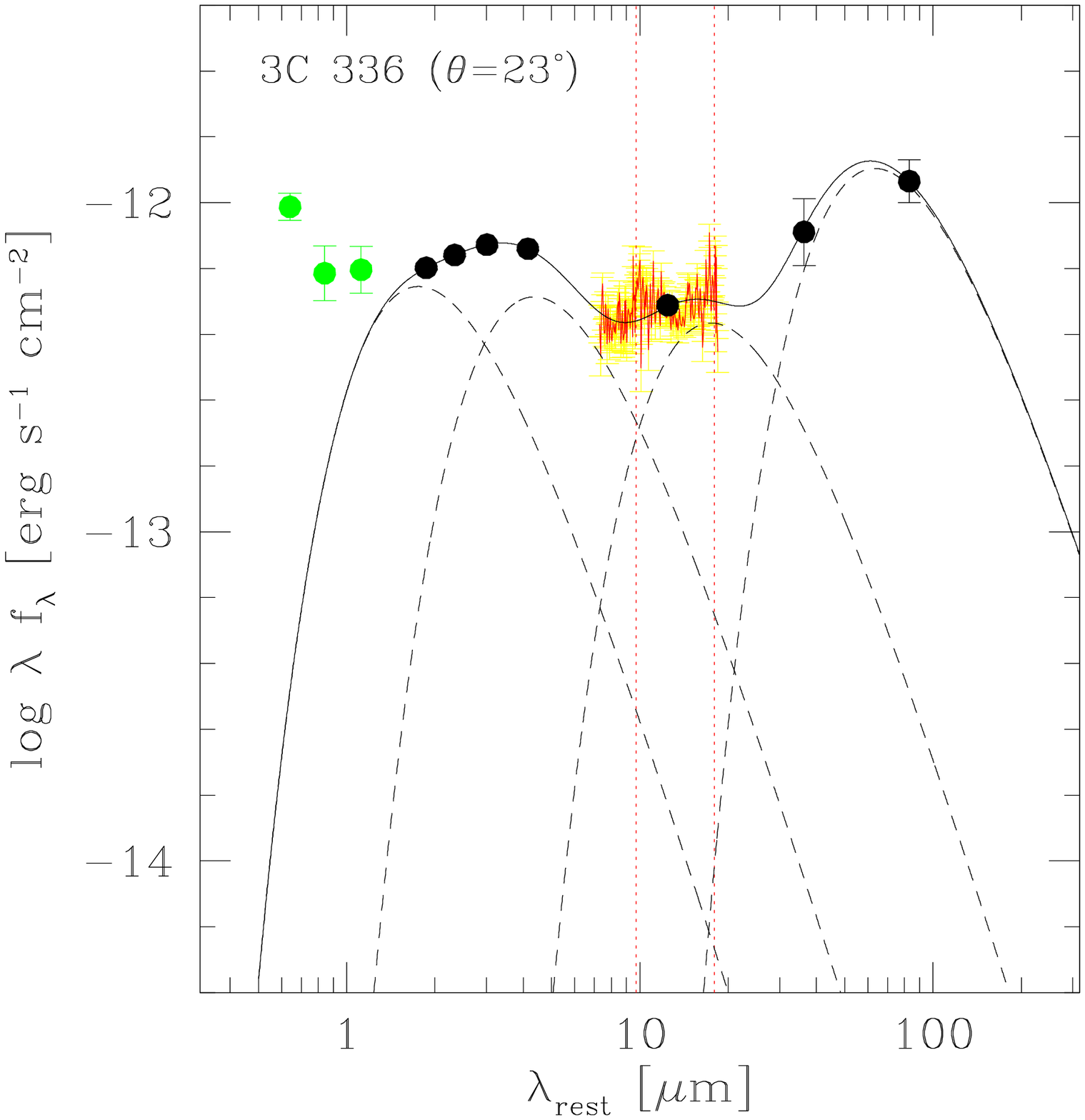}
\includegraphics[scale=0.28]{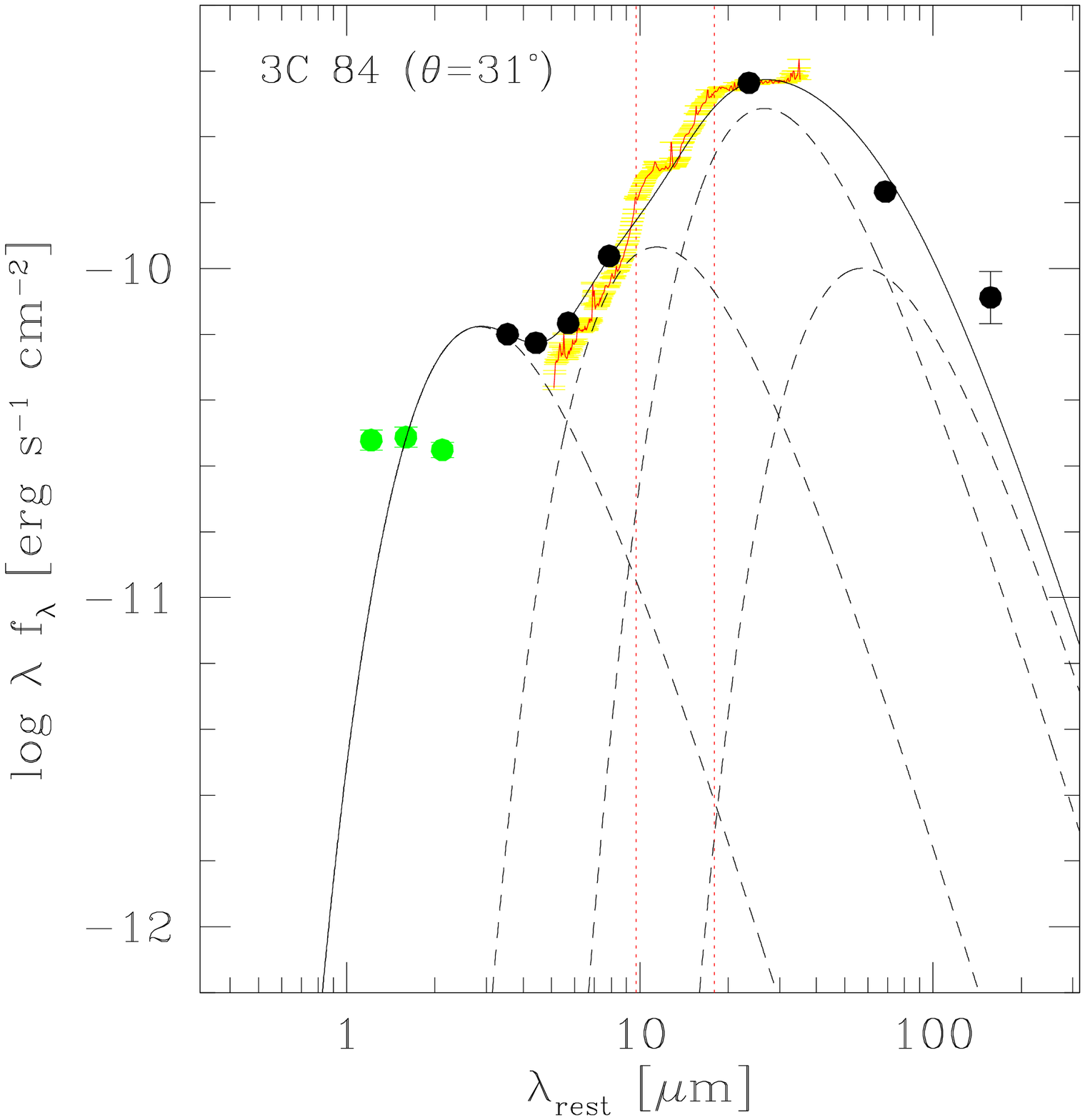}
\includegraphics[scale=0.28]{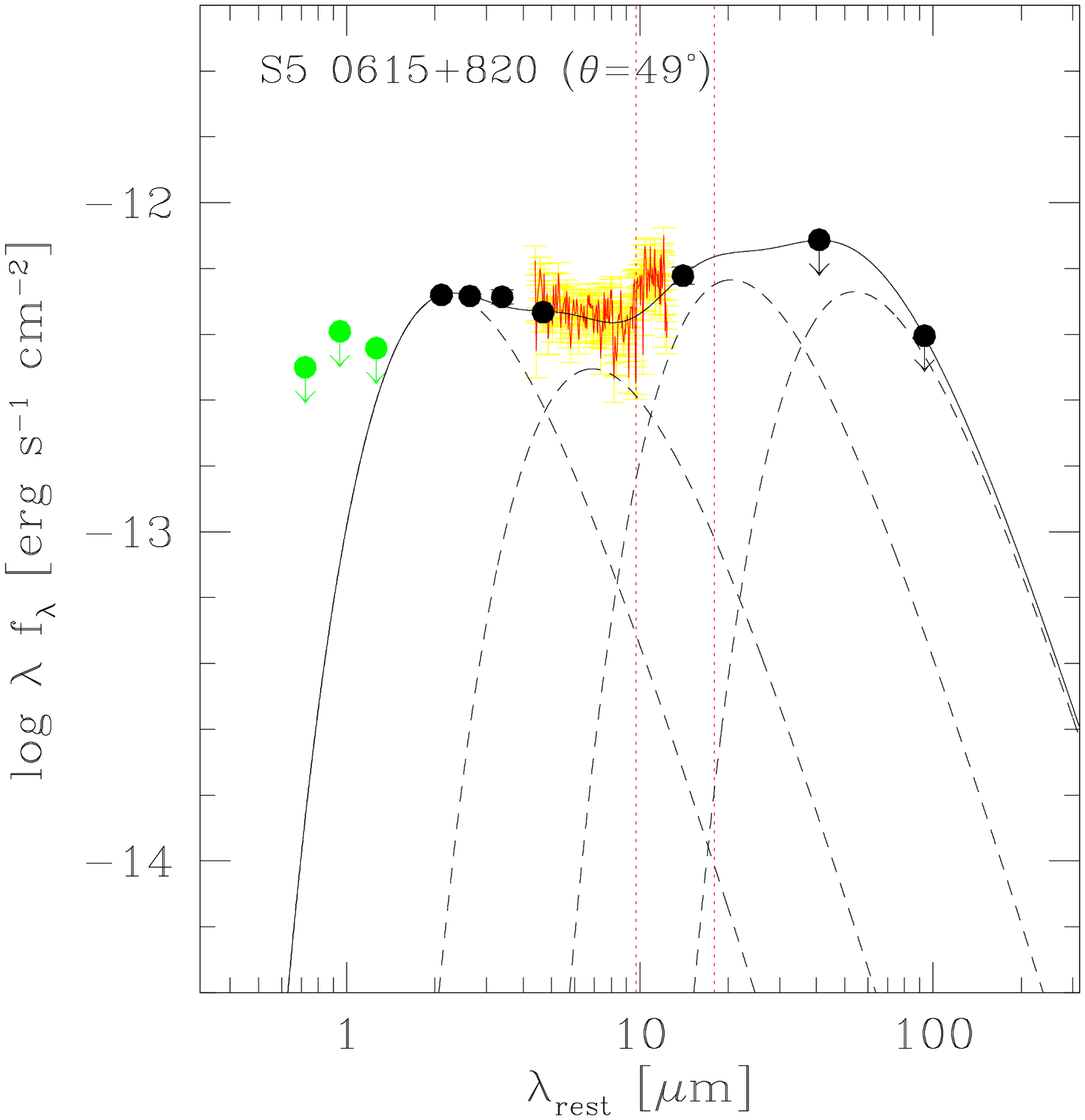}
}
\caption{\label{sed} Infrared spectral energy distributions plotted as
  rest-frame wavelength versus observed integrated flux. Filled black
  and green circles indicate {\it Spitzer} IRAC and MIPS and 2MASS
  photometry, respectively. The {\it Spitzer} IRS spectrum is shown in
  red, with its errors in yellow and the locations of the silicate
  features marked (red dotted lines). The dotted and solid black
  curves show the best-fit blackbodies and their sum,
  respectively. The fit for the source 3C 390.3 included the IRAC data
  scaled to the IRS spectrum (open black circles).}
\end{figure*}

Fig. \ref{sed} shows the infrared SEDs for our sources plotted to have
the same dynamic range. In the near-infrared we have supplemented the
{\it Spitzer} data with ground-based photometry in the $J$, $H$ and
$K_s$ bands from the Two Micron All-Sky Survey
\citep[2MASS;][]{2MASS}. All sources but one are included in the Point
Source Catalogue of this survey and we list the flux densities in
Table \ref{suppl}. For the undetected source S5~0615$+$820 we have
derived $3 \sigma$ upper limits based on the image plate
specifications.


\subsection{General trends}

Before testing detailed theoretical models with our data (Section
\ref{models}) we want to first quantify the observations, with
particular emphasis on revealing trends that any acceptable model will
have to account for.

\subsubsection{The shape and width of the SED} \label{blackbody}


\begin{table*}
\caption{\label{dust} Blackbody Fit Results}
\begin{tabular}{lcccccccccc}
\hline
Object Name$^{\star}$ & $\theta$ & 
$T_{\rm hot}$ & $T_{\rm warm}$ & $T_{\rm cool}$ & $T_{\rm cold}$ & 
$f_{\rm hot}$ & $f_{\rm warm}$ & $f_{\rm cool}$ & $f_{\rm cold}$ & $\chi_{\nu}^2$/dof \\
& [deg] & [K] & [K] & [K] & [K] & [erg/s/cm$^2$] & [erg/s/cm$^2$] & [erg/s/cm$^2$] & [erg/s/cm$^2$] \\
(1) & (2) & (3) & (4) & (5) & (6) & (7) & (8) & (9) & (10) & (11) \\
\hline
3C 207        &  8 & 1459 & 509 & 133 &  -- & 1.33e$-$12 & 1.05e$-$12 & 2.46e$-$12 & --         & 5.3/8  \\
3C 47         & 18 & 1251 & 456 & 149 &  -- & 4.48e$-$12 & 2.20e$-$12 & 4.58e$-$12 & --         & 1.62/7 \\
3C 245        & 18 & 1543 & 501 & 139 &  -- & 1.47e$-$12 & 2.78e$-$12 & 2.95e$-$12 & --         & 8.1/98 \\
3C 336        & 23 & 2087 & 842 & 210 &  58 & 5.56e$-$13 & 5.17e$-$13 & 4.31e$-$13 & 1.27e$-$12 & 1.27/4 \\
4C $+$28.45   & 24 & 1175 & 403 & 146 &  -- & 2.27e$-$12 & 8.92e$-$13 & 2.78e$-$12 & --         & 1.99/7 \\
III Zw 2      & 28 & 1561 & 514 & 176 &  69 & 1.87e$-$11 & 1.37e$-$11 & 1.37e$-$11 & 2.63e$-$12 & 3.8/7  \\
4C $+$34.47   & 30 & 1702 & 632 & 189 &  -- & 6.80e$-$12 & 5.27e$-$12 & 6.53e$-$12 & --         & 4.2/7  \\
3C 84         & 31 & 1290 & 321 & 138 &  64 & 6.65e$-$11 & 1.16e$-$10 & 3.07e$-$10 & 1.00e$-$10 & 1491/8 \\
3C 334        & 34 & 1733 & 799 & 207 &  98 & 2.31e$-$12 & 1.85e$-$12 & 4.00e$-$12 & 1.76e$-$12 & 3.7/7  \\
3C 263        & 48 & 1581 & 666 & 247 & 133 & 4.37e$-$12 & 2.27e$-$12 & 2.04e$-$12 & 1.80e$-$12 & 1.42/7 \\
3C 390.3$^\dagger$&48&1274& 488 & 183 &  -- & 3.04e$-$11 & 1.08e$-$11 & 2.53e$-$11 & --         & 757/11 \\
S5 0615$+$820 & 49 & 1654 & 533 & 179 &  68 & 5.22e$-$13 & 3.13e$-$13 & 5.83e$-$13 & 5.36e$-$13 & 0.55/6 \\
\hline
\end{tabular}

\medskip

\parbox[]{16.5cm}{The columns are: (1) object name; (2) jet inclination
  angle; temperature of the (3) hot, (4) warm, (5) cool, and (6) cold
  blackbody component; peak flux of the (7) hot, (8) warm, (9) cool,
  and (10) cold blackbody component; and (11) reduced $\chi^2$ value and
  number of degrees of freedom.}

\medskip

\parbox[]{16.5cm}{$^{\star}$ objects presented in order of increasing jet viewing angle}
\parbox[]{16.5cm}{$^{\dagger}$ fit performed to the IRAC data scaled to the IRS spectrum}

\end{table*}


\begin{figure*}
\centerline{ 
\includegraphics[scale=0.3]{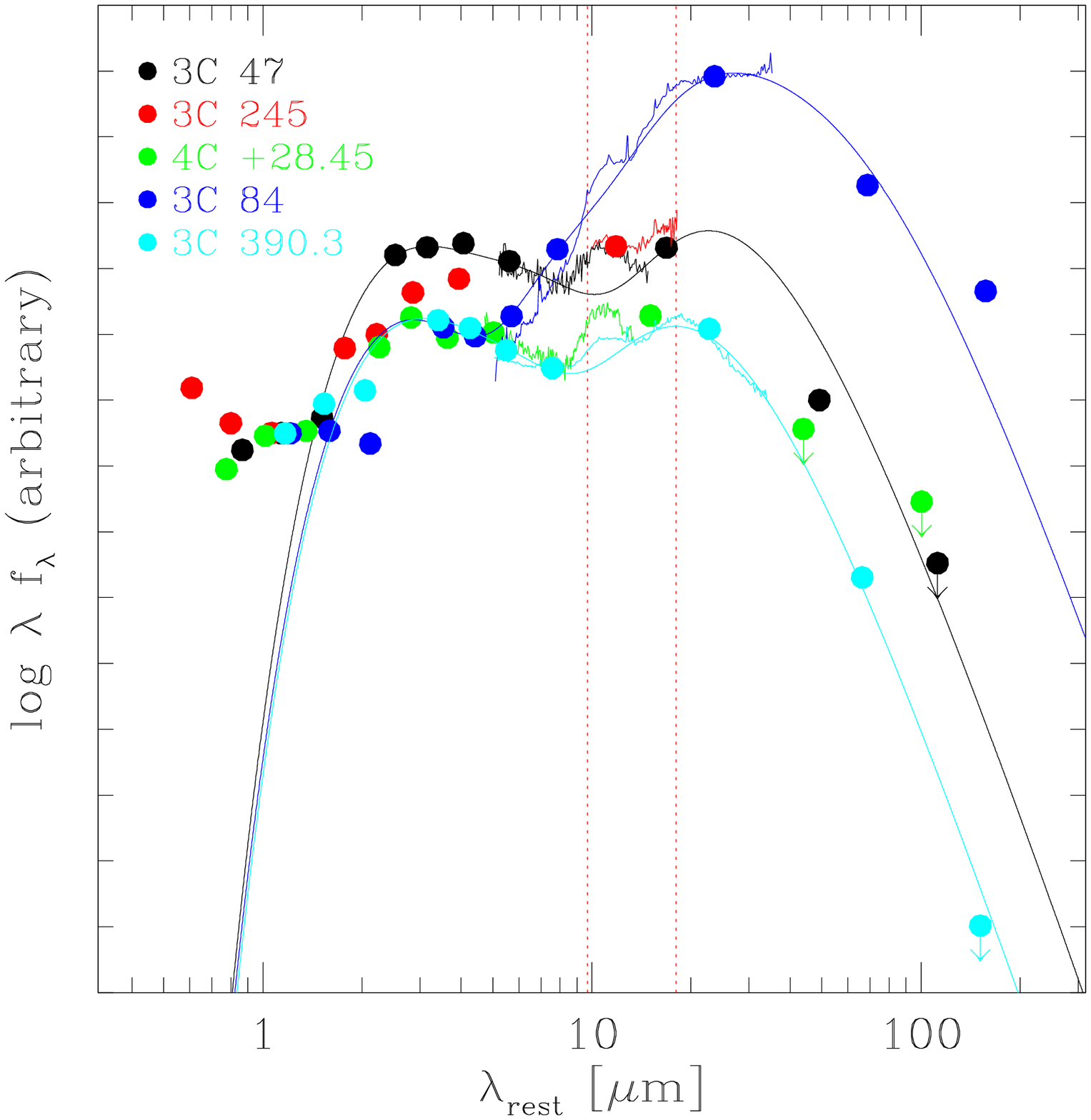}
\includegraphics[scale=0.3]{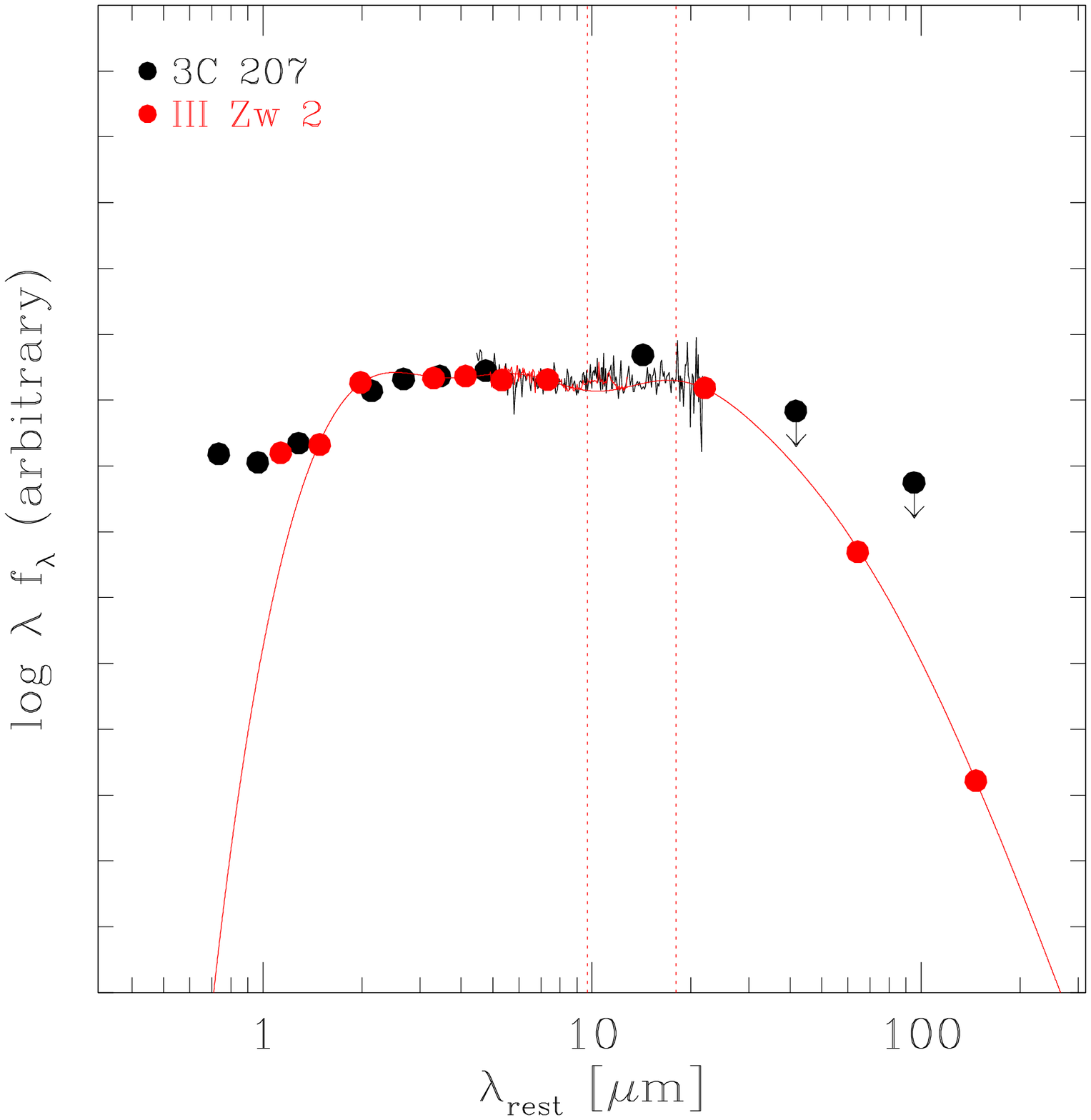}
\includegraphics[scale=0.3]{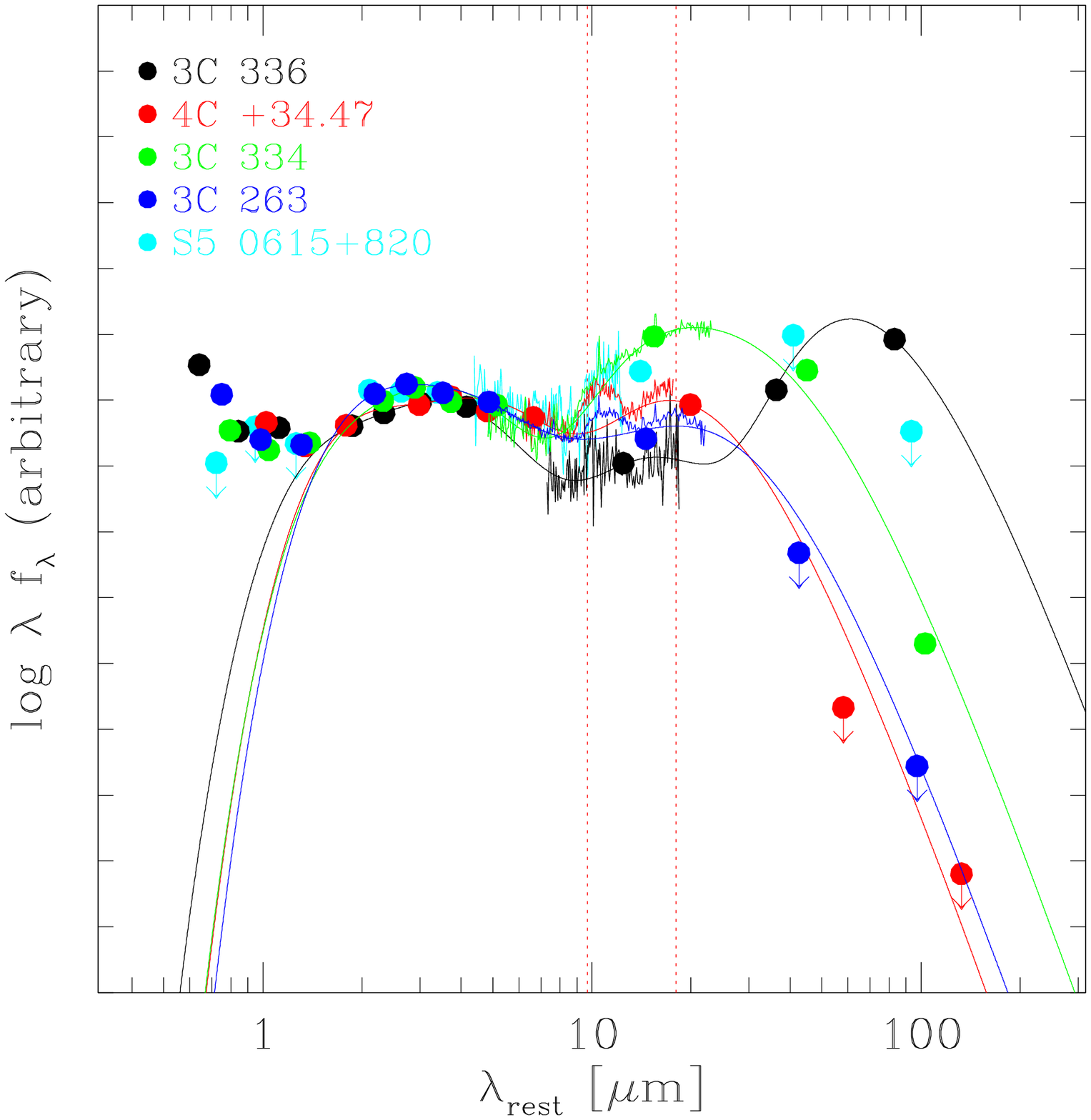}
}
\caption{\label{sedtotal} Infrared spectral energy distributions from
  Fig. \ref{sed} normalized at $\sim 1$~$\mu$m, grouped by the
  relative strength of the hot blackbody component and shown from left
  to right in decreasing order. The solid line indicates the sum of
  the best-fit blackbodies and has been omitted for the sources
  3C~207, 3C~245, 4C~$+$28.45, and S5~0615$+$820, for which the
  long-wavelength end is not well constrained. The locations of the
  silicate features are marked by the red dotted lines.}
\end{figure*}

In the simplest approach we can approximate our observations with a
set of blackbodies. For this purpose we have fitted our data with the
C routine MPFIT \citep[version 1.1;][]{mpfit}, which uses the
Levenberg-Marquardt technique to solve the least-squares problem. We
have tried several numbers of blackbodies and have found that three or
four components were required to obtain an acceptable fit. We have
fitted for the temperatures and flux scalings of the individual
blackbody components, meaning that our fits have either six or eight
free parameters. Our results are listed in Table \ref{dust} and shown
in Fig. \ref{sed}, where dotted black curves indicate the individual
components and the solid black curve their sum.

We have included in the fit only the continuum part of the IRS
spectrum, i.e., we have excluded strong narrow emission lines and the
silicate emission features, and we have rebinned it to $\Delta \log
\lambda = 0.05~\mu$m in order to ensure a similar weighting between
spectroscopy and photometry. However, in the case of the source 3C
245, for which neither MIPS-70 nor MIPS-160 data were available, we
have left the spectrum unbinned. We have treated all photometry upper
limits as detections and have assumed their 1$\sigma$ values as the
error. We have not considered the 2MASS photometric data points, since
not only they appear to sample a different component (most likely the
onset of the accretion disc) but they were obtained several years
before our {\it Spitzer} observations. The two exceptions were III Zw
2 and 4C $+$34.47, for which we included the 2MASS $K_{\rm s}$ point,
since it connected smoothly to the IRAC photometry and thus presented
an important constraint on the hottest blackbody component. The IRAC
data we use for the source 3C 390.3 was taken four years after the
MIPS and IRS observations and variability by a factor of $\sim 1.5$ is
observed. Therefore, assuming that variability does not change the
spectral slope, we have scaled these data to the IRS spectrum before
including them in the fit.

Our sample is evenly split into sources best-fit by three and four
blackbodies. For the source 3C 245, the far-infrared part of the SED
is not constrained and, therefore, we cannot exclude that four instead
of three blackbodies might be required. In all sources but two the
resulting fit is `good' in a statistical sense ($\chi_{\nu}^2$ of a
few). In the two brightest sources (3C 84 and 3C 390.3) the resulting
fit appears good to the eye and the large $\chi_{\nu}^2$-values could
be mainly due to the much smaller relative measurement errors
involved. Resulting temperatures for the hot, warm, cold and cool
blackbody components are in the ranges $T_{\rm hot} \sim 1200 - 2000$
K, $T_{\rm warm} \sim 300 - 800$ K, $T_{\rm cool} \sim 150 - 250$ K,
and $T_{\rm cold} \sim 60 - 150$ K, respectively. Note that the
hottest blackbody component reaches values that are typical of the
dust sublimation temperature for most grain compositions
\citep[$\approx 1000 - 2000$ K;][]{Sal77}.

Three important trends are revealed by this simplistic
approach. Firstly, the strengths of the individual blackbodies
relative to each other vary substantially between sources, giving the
impression that they indeed sample discrete components of a certain
temperature rather than a single component with a smooth temperature
distribution. Secondly, the cool ($\sim 200$ K) blackbody component is
prominent in all sources and appears roughly as strong or stronger
than the hot blackbody component. Only in the source 3C~263 is this
behaviour reversed. And thirdly, the warm ($\sim 500$ K) blackbody
component is often weak, thus introducing a sharp 'dip' in the SED
just blueward of the 10 $\mu$m silicate feature. We also note that,
although in most cases we did not include the 2MASS $K_{\rm s}$ point
in the fit, it is often approximated well by the hot blackbody
component.

In order to investigate if orientation determines to some degree the
relative strengths of the individual blackbody components in a given
source, we have normalized the SEDs at $\sim 1$~$\mu$m and have
grouped them in order of the relative strength of the hot blackbody
component. In Fig. \ref{sedtotal} we show our sources in three
distinct groups with the average relative strength of the hot
blackbody component decreasing from left to right. This representation
of the SEDs suggests two important results. Firstly, no trend with
orientation seems to be present for the hot dust. In particular,
although the two sources with the strongest hot blackbody components
(3C~47 and 3C~245) are among those with the smallest estimated viewing
angles, the main prediction of torus models assuming continuous
density distributions, namely that the relative emission from hot dust
increases with decreasing viewing angle \citep[see, e.g.,][their
Figs. 4 and 5]{Pier92}, is in general not observed for our sample.

\begin{figure}
\centerline{ 
\includegraphics[scale=0.4]{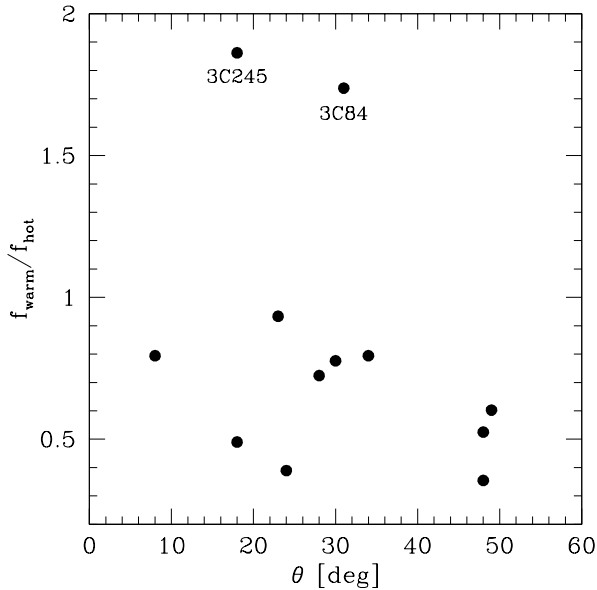}
}
\caption{\label{warmangle} The ratio between the peak fluxes of the
  warm and hot blackbody components versus the jet viewing angle.}
\end{figure}

\begin{figure}
\centerline{
\includegraphics[scale=0.4]{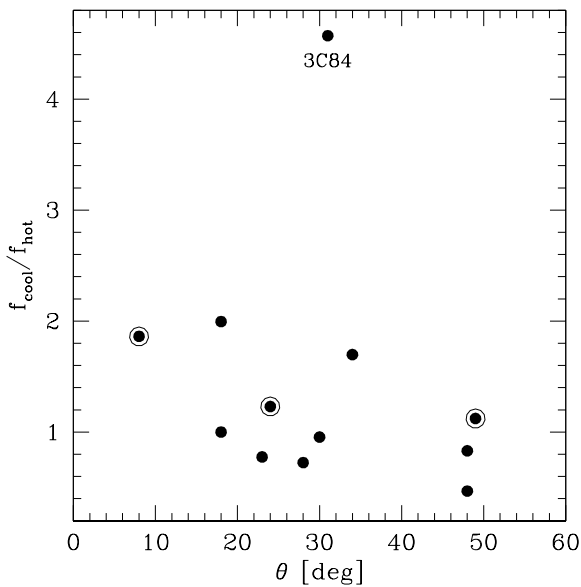}
}
\caption{\label{coolangle} Same as Fig. \ref{warmangle} for the cool
  blackbody component. This component is likely to be overestimated
  for three sources (3C~207, 4C~$+$28.45, and S5~0615$+$820; encircled
  points).}
\end{figure}

Secondly, it appears that at any viewing angle the SEDs can have
widely different shapes. However, we find trends with orientation for
the relative strengths of both the warm and cool dust
components. These trends, which are illustrated in
Figs. \ref{warmangle} and \ref{coolangle}, respectively, suggest that
as the viewing angle decreases the warm and cool dust emissions
relative to that of the hot component {\it increase}. This result is
contrary to the expectations of smooth-density torus models
\citep[see, e.g.,][their Figs. 5 and 7]{Pier92}. The source 3C~84
stands out in both Figs. \ref{warmangle} and \ref{coolangle} as having
a relatively weak hot dust component.

With the help of Fig. \ref{sedtotal} we can identify differences and
similarities between sources. Of special interest are possible
shortcomings in the fits for those five sources that have two upper
limits at the long-wavelength end, namely, 3C~207, 4C~$+$28.45,
4C~$+$34.47, 3C~263, and S5~0615$+$820, and the poorly constrained
source 3C~245. Firstly, we note that the values of the upper limits
for the two sources 4C~$+$34.47 and 3C~263 are relatively low and,
although located at different rest-frame wavelengths, appear to sample
a similar cool blackbody component. Therefore, they are unlikely to be
far from the true values. Secondly, based on the similarity between
the SEDs of the sources 3C~207, 4C~$+$28.45 and S5~0615$+$820, and
those of the sources III~Zw~2, 3C~390.3, and 4C~$+$34.47,
respectively, the peak of the cool blackbody component of the former
is unlikely to be overestimated by factors $\ga2$. Finally, judging
from a comparison between the SEDs of the sources 3C~245 and 3C~47,
whereas the peaks of their cool blackbody components appear similar,
the peak of the warm blackbody component of the former could be
overestimated by a factor of $\sim 2$.

\subsubsection{The silicate features} \label{sifeature}


\begin{table*}
\caption{\label{silicates} 
Properties of the Silicate Emission Features}
\begin{tabular}{lccccccccc}
\hline
Object Name$^\star$ & $\theta$ & \multicolumn{4}{c}{Silicate 10 $\mu$m} & \multicolumn{4}{c}{Silicate 18 $\mu$m} \\
&& flux & luminosity & W$_{\lambda}$ & center & flux & luminosity & W$_{\lambda}$ & center \\
&& [erg/s/cm$^2$] & [erg/s] & [$\mu$m] & [$\mu$m] & [erg/s/cm$^2$] & [erg/s] & [$\mu$m] \\
(1) & (2) & (3) & (4) & (5) & (6) & (7) & (8) & (9) & (10) \\
\hline
3C 207        &  8 & 4.42e$-$14 & 8.91e$+$43 & 0.65 & 10.31 & --         & --         & --   & --    \\
3C 47         & 18 & 2.87e$-$13 & 1.86e$+$44 & 1.30 & 10.65 & ?          & ?          & ?    & ?     \\
3C 336        & 23 & 8.45e$-$15 & 3.63e$+$43 & 0.37 & 10.13 & --         & --         & --   & --    \\
4C $+$28.45   & 24 & 1.30e$-$13 & 1.91e$+$44 & 1.34 & 10.61 & ?          & ?          & ?    & ?     \\
III Zw 2      & 28 & 2.38e$-$13 & 4.79e$+$42 & 0.16 & 10.53 & ?          & ?          & ?    & ?     \\
4C $+$34.47   & 30 & 4.66e$-$13 & 5.75e$+$43 & 0.93 & 10.44 &$>$1.04e$-$13&$>$1.29e$+$43&$>$0.29& ?  \\
3C 84         & 31 & 6.83e$-$12 & 5.01e$+$42 & 0.47 & 10.83 & 1.02e$-$11 & 7.41e$+$42 & 0.61 & 16.95 \\
3C 334        & 34 & 7.94e$-$14 & 9.77e$+$43 & 0.43 & 10.53 & --         & --         & --   & --    \\
3C 263        & 48 & 1.05e$-$13 & 1.86e$+$44 & 0.56 & 10.92 & 4.31e$-$14 & 7.76e$+$43 & 0.34 & 17.64 \\
3C 390.3$^\dagger$&48&1.22e$-$12& 9.12e$+$42 & 0.59 & 10.37 & 4.13e$-$13 & 3.09e$+$42 & 0.27 & 18.39 \\
S5 0615$+$820 & 49 &$>$1.11e$-$14&$>$2.51e$+$43&$>$0.42&$>$10.61& ?          & ?          & ?    & ?     \\
\hline
\end{tabular}

\medskip

\parbox[]{15.5cm}{The columns are: (1) object name; (2) jet viewing
  angle; for the 10 $\mu$m silicate emission feature (3) integrated
  flux, (4) luminosity, (5) rest-frame equivalent width, and (6)
  rest-frame center of the emission; for the 18 $\mu$m silicate
  emission feature (7) integrated flux, (8) luminosity, (9) rest-frame
  equivalent width and (10) rest-frame center of the emission. A
  question mark indicates that the feature is not covered by the
  spectrum.}

\medskip

\parbox[]{15.5cm}{$^{\star}$ objects presented in order of increasing jet viewing angle}
\parbox[]{15.5cm}{$^{\dagger}$ continuum fit used the IRAC data scaled to the IRS spectrum}

\end{table*}


A characteristic property of interstellar dust are the spectral
features centered at 9.7 $\mu$m and 18 $\mu$m due to silicates. In
AGN, these features are expected to be produced by the dusty torus. A
pertinent problem inherent to the measurement of silicate features
remains the correct placement of the continuum, which ultimately
decides if they are seen in emission or in absorption. This task is
still challenging because the available spectra rarely cover a
wavelength region large enough to see a sizeable portion of the
continuum around the features, which are relatively broad and often
shifted in wavelength (see Fig. \ref{sed}). The advantage of our data
set, however, is that the IRAC and MIPS photometry considerably extend
the wavelength range of the IRS spectrum, thus allowing us to detect
the {\it overall} continuum. In particular, our data shows that the
'dip' in the SED blueward of $\sim$ 10 $\mu$m observed in most
sources, which could be interpreted as blueshifted silicate
absorption, is in fact the result of a warm dust component that is
relatively weak.

Based on the overall continuum, we detect silicate {\it emission} in
all our sources. The only exception is the source 3C~245, for which we
do not detect strong silicate emission and interpret the IRS spectrum
as being dominated by the warm and cool dust components. We have
measured the properties of both the 10 $\mu$m and 18 $\mu$m silicate
emission features after subtracting from the IRS spectrum the overall
continuum fitted in Section \ref{blackbody} and removing superimposed
strong narrow emission lines. Our results are listed in Table
\ref{silicates}. We detect the 18 $\mu$m silicate feature in 4/8
sources, for which the IRS spectrum covers its location. In the
remaining four sources this feature appears to be swamped by the cool
blackbody component, which, based on its temperature ($\sim 200$ K),
peaks around this wavelength.

Our first noteworthy finding is that the 10 $\mu$m silicate emission
feature is considerably redshifted in all our sources. Instead of at
the expected rest-frame wavelength of $\sim 9.7~\mu$m, we find its
center to lie in the range of $\sim 10.1 - 10.9~\mu$m (Table
\ref{silicates}, column (6)). We note that strongly redshifted
silicate emission features were observed previously in a few quasars
\citep{Sieben05, Hao05, Schweitzer08}, but never consistently in an
entire sample. On the other hand, the center of the 18 $\mu$m silicate
emission feature in the three objects for which it can be determined
is strongly shifted in only one source (3C~84) and in this case
blueward.

\begin{figure}
\centerline{ 
\includegraphics[scale=0.4]{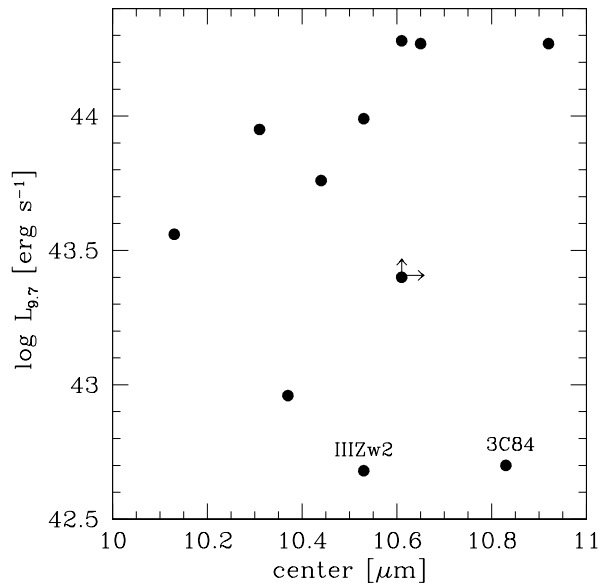}
}
\caption{\label{silum} The luminosity of the 10 $\mu$m silicate
  emission feature versus the feature center. Arrows indicate limits.}
\end{figure}

From circumstellar dust studies we know that the exact wavelength
location of the silicate emission maximum depends on the grain size
and composition, with larger grain sizes and larger amounts of
crystalline dust (over amorphous dust) expected to enhance the
emissivity at longer wavelengths \citep[e.g.,][]{Bou01}. In order to
investigate if this assumption holds also for AGN, we have plotted in
Fig. \ref{silum} the luminosity of the 10 $\mu$m silicate feature
versus its rest-frame center. Indeed, we find that stronger silicate
emission is associated with a stronger redward displacement. However,
the two lowest-luminosity sources (III~Zw~2 and 3C~84) do not appear
to follow this trend.

We note that in the current literature authors often study the
(emission) equivalent widths of the silicate features rather than
their luminosities. This approach needs to be taken with caution,
since, by definition, the equivalent width depends strongly on the
continuum flux. And, as the SEDs in Fig. \ref{sed} show, the 10 $\mu$m
silicate feature is located such that its continuum flux is given by
the relative strengths of the warm and cool blackbody components,
which vary strongly between sources. In fact, the relation observed in
Fig. \ref{silum} is not evident if we use rest-frame equivalent width
values instead of luminosities, and our sample shows a trend for
higher equivalent widths to be measured, the weaker the warm blackbody
component is relative to the cool one.

\subsubsection{The jet emission} \label{jet}

\begin{figure}
\centerline{
\includegraphics[scale=0.4]{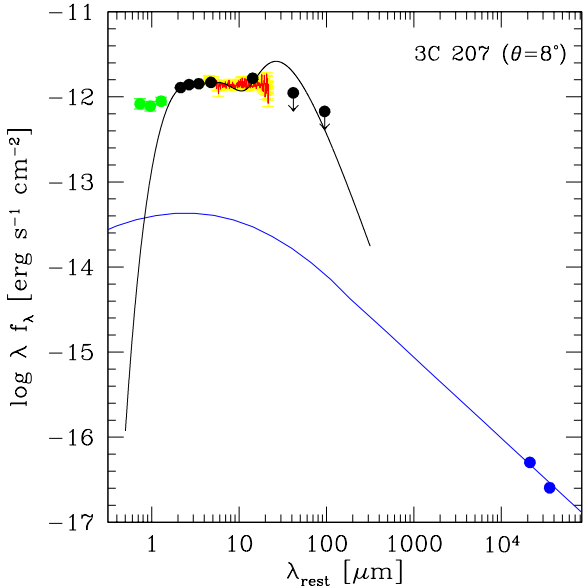} 
}
\caption{\label{jetsed} Infrared spectral energy distribution for 3C
  207 from Fig. \ref{sed} extended to radio frequencies. An
  extrapolation of the radio core spectrum [8 GHz VLBI data
  \citep{Hough02} and 5 GHz VLA data \citep{Har04}; blue filled
  circles] to higher frequencies using a typical blazar SED (blue
  solid curve) shows that even in our strongest relativistically
  beamed source the jet does not dominate the infrared emission.}
\end{figure}

\begin{figure}
\centerline{
\includegraphics[scale=0.4]{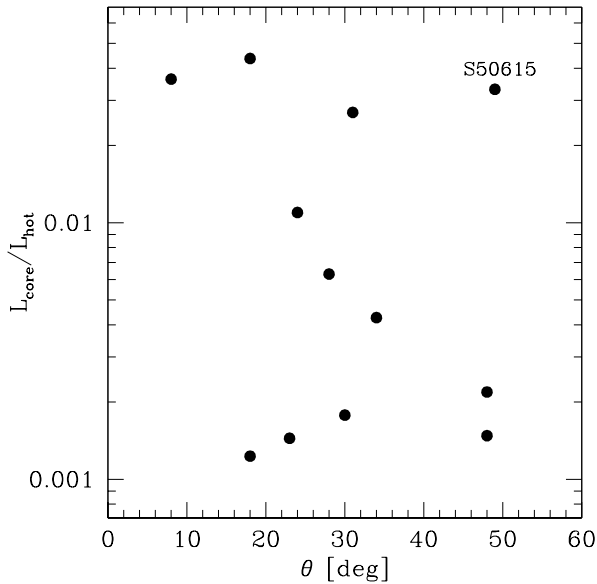}
}
\caption{\label{vatest} The ratio between the integrated core radio
  luminosity and the peak luminosity of the hot blackbody component
  versus the jet viewing angle.}
\end{figure}

We have selected our sample of radio-loud AGN based on their
relatively large jet viewing angles ($\theta \ga 20^{\circ}$). At
these orientations the relativistic enhancement of the jet emission is
expected to be very low, ensuring that the observed infrared SED is
dominated by emission from the putative dusty torus. In fact, based on
the calculated relativistic Doppler factors $\delta$ (Table
\ref{general}, column (16)), we expect relativistic beaming of the
integrated jet flux, which is proportional to $\delta^4$, by factors
of $\la 10$ in three sources (III~Zw~2, 3C~84, and 3C~245) and
relativistic {\it debeaming} for the remainder.

The exception is the source 3C~207, for which, based on improved
proper motion data, we now calculate a relatively small inclination
angle ($\theta \sim 8^{\circ}$). Its relativistic Doppler factor is
$\delta \sim 3$, meaning that the relativistic jet flux enhancement is
expected to be a factor of $\sim 80$. However, as Fig. \ref{jetsed}
shows, even in the strongest relativistically beamed source in our
sample the infrared SED is dominated by thermal emission. An
extrapolation of the core radio spectrum using a typical blazar SED
that generally peaks at a few $\mu$m \citep{Gio02} predicts
mid-infrared fluxes a factor of $\sim 40$ lower than observed.

Further evidence that the infrared SEDs of all our sources are
dominated by thermal rather than non-thermal emission comes from their
similar, relatively bulgy appearance and the lack of strong,
short-term variability. The IRS spectroscopy overlaps in wavelength
with either the IRAC and/or the MIPS photometry, and an impressive
consistency is evident between the two data sets, which were taken
several months apart (see Tables \ref{spitzerphot} and
\ref{spitzerspec}). Variability is detected in the source 3C~390.3,
however, not of the strength typical of blazars. The flux increase is
a factor of $<2$ over a period of four years.

Given that the infrared emission in our sources is thermal and that
the hot dust emission seems to be emitted isotropically (see Section
\ref{blackbody}), we can now carry out a consistency check for the
calculated viewing angles. Along the lines of argument presented by
\citet{Wills95}, we plot in Fig. \ref{vatest} the ratio between the
integrated core radio luminosity (calculated from the data listed in
Table \ref{general}) and the peak luminosity of the hot blackbody
component (calculated from the fluxes listed in Table \ref{dust},
column (7)) versus the jet viewing angle. In such a diagram, we expect
that the stronger a source is relativistically beamed, the higher its
ratio between beamed and isotropic emission. Fig. \ref{vatest} shows
that the resulting trend in viewing angle for our sample is
qualitatively correct; the smaller the viewing angle, the higher the
ratio between (beamed) core radio power and (isotropic) dust
luminosity. The only pronounced exception is the source S5~0615$+$820,
for which the viewing angle appears to be highly overestimated.

\subsection{Testing CLUMPY torus models} \label{models}

\begin{figure}
\centerline{ 
\includegraphics[scale=0.4]{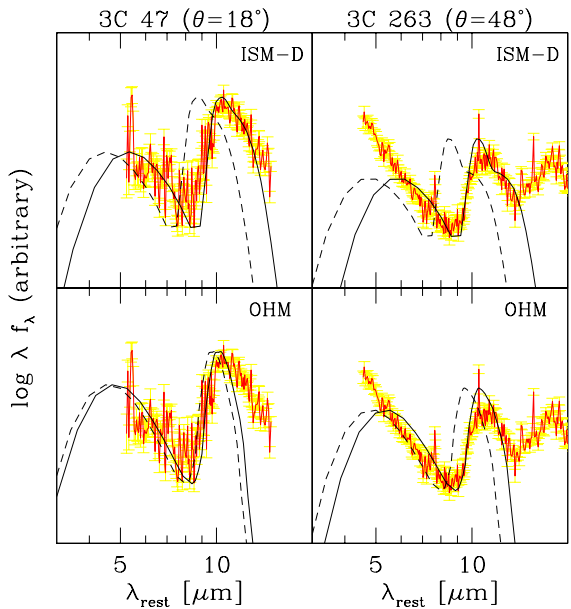}
}
\caption{\label{clumpydust} The 10 $\mu$m silicate emission feature
  and the spectral slope blueward of it are well approximated by {\it
    redshifted} CLUMPY models with standard ISM dust composition (code
  ISM-D; top panels, solid lines). Dust composed of silicates with
  various materials added (code OHM; bottom panels) does not reproduce
  the observations as well, independent of whether we apply a redshift
  (solid lines) or not (dashed lines). }
\end{figure}

\begin{figure}
\centerline{ 
\includegraphics[scale=0.4]{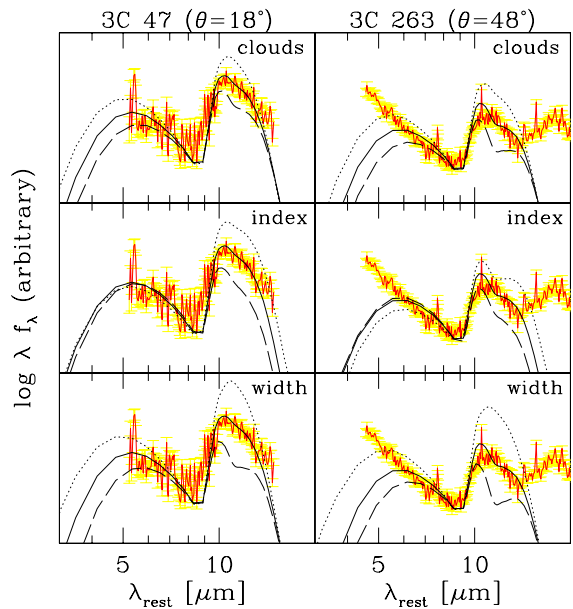}
}
\caption{\label{clumpydens} The height of the 10 $\mu$m silicate
  emission peak relative to the 'dip' blueward of it constrains best
  the number of clouds (top panels, where dotted, solid and dashed
  lines indicate $N_0=4$, 5 and 6, respectively), the power-law index
  of the radial cloud distribution (middle panels, where dotted, solid
  and dashed lines indicate $q=1$, 2 and 3, respectively), and the
  angular width of the Gaussian cloud distribution (bottom panels,
  where dotted, solid and dashed lines indicate $\sigma=30^\circ$,
  45$^\circ$ and 60$^\circ$, respectively).}
\end{figure}

Due to the shortcomings in AGN torus models invoking continuous dust
distributions, new models based on clumpy media have recently been put
forward \citep[e.g.,][]{Schart05, Schart08, Hoenig06, Hoenig10b,
  Nen02, Nen08a, Nen08b}. The most detailed of these models are those
of Nenkova and collaborators (named CLUMPY), and they can be accessed
on-line\footnote{See \url{https://newton.pa.uky.edu/~clumpyweb/}}. In
short, these authors solved the radiative transfer problem in clumpy
media by assuming that the medium is composed of clouds that are
individually optically thick ($\tau_{\rm V}$$\gg$1), that each cloud
can be considered a point source of intensity $S_{\rm c,\lambda}$, and
that the cloud distribution obeys Poisson statistics. In this case,
the escape probability of the emitted radiation can be approximated as
$P_{\rm esc} \simeq e^{-N_{\rm T}}$, with $N_{\rm T}$ the total number
of clouds along the line of sight, and the intensity at a given
location becomes:

\begin{equation}
\label{inteq}
I_{\lambda}^{\rm C} (s)  = \int\limits^{s} e^{-N_{\rm T}(s,s^\prime)} S_{\rm c,\lambda} N_C(s^\prime) ds^\prime,
\end{equation}

\noindent
where $N_{\rm C}$ is the radial cloud density (i.e., the number of
clouds per unit length). With this formalism the only difference
between the clumpy and continuous cases is that optical depth
$\tau_{\rm V}$ is replaced by its effective equivalent $N_{\rm
  T}(1-e^{-\tau_{\rm V}}) \simeq N_{\rm T}$ (for $\tau_{\rm V}$$\gg$1)
and the absorption coefficient is replaced by $N_{\rm C}$. The main
challenge lies with the calculation of the clump source function,
$S_{\rm c,\lambda}$, that needs to properly take into account the
illumination profile of the individual clouds and the effects of cloud
shadowing.

In this section we wish to test the CLUMPY models, in particular how
well they approximate the entire infrared SED, how unique a fitting
set of parameters is and what physics can be extracted. For
simplicity, we will limit ourselves to models invoking a torus
geometry with a Gaussian angular cloud distribution, i.e., we will not
consider a spherical distribution or the sharp-edge geometry.

\subsubsection{Constraining the parameters}


\begin{table*}
\caption{\label{clumpypar} 
CLUMPY Torus Model Fit Results}
\begin{tabular}{lcccccccc}
\hline
Object Name$^{\star}$ & $\theta$ & $z_{\rm C}$ & $\tau_{\rm V}$ & $N_0$ & $q$ & $L_{\rm bol}$ & $R_{\rm in}$ & $M_{\rm C}$ \\
& [deg] &&&&& [erg/s] & [pc] & [M$_\odot$] \\
(1) & (2) & (3) & (4) & (5) & (6) & (7) & (8) & (9) \\
\hline
3C 207        &  8 & 0.225 &  10 &  7 & 3 & 1.12e$+$46 & 1.340 & 8.5e$+$04 \\
3C 47         & 18 & 0.186 &  10 &  5 & 2 & 1.15e$+$46 & 1.355 & 1.3e$+$05 \\
3C 245        & 18 & 0.225 & 150 & 20 & 3 & 3.47e$+$46 & 2.355 & 1.1e$+$07 \\
3C 336        & 23 & 0.202 &  10 &  5 & 3 & 9.77e$+$45 & 1.250 & 5.3e$+$04 \\
4C $+$28.45   & 24 & 0.167 &  10 &  4 & 2 & 1.41e$+$46 & 1.503 & 1.3e$+$05 \\
III Zw 2      & 28 & 0.230 &  10 &  7 & 3 & 1.48e$+$45 & 0.486 & 1.1e$+$04 \\
4C $+$34.47   & 30 & 0.186 &  10 &  5 & 2 & 4.07e$+$45 & 0.807 & 4.7e$+$04 \\
3C 84         & 31 & 0.109 & 150 & 20 & 3 & 3.09e$+$44 & 0.222 & 1.0e$+$05 \\
3C 334        & 34 & 0.067 & 150 & 20 & 3 & 9.33e$+$45 & 1.222 & 3.0e$+$06 \\
3C 263        & 48 & 0.230 &  10 &  5 & 2 & 3.55e$+$46 & 2.383 & 4.1e$+$05 \\
3C 390.3      & 48 & 0.175 &  10 &  5 & 1 & 9.55e$+$44 & 0.391 & 2.4e$+$04 \\
S5 0615$+$820 & 49 & 0.202 &  10 &  4 & 1 & 6.61e$+$45 & 1.028 & 1.3e$+$05 \\
\hline
\end{tabular}

\medskip

\parbox[]{11cm}{The columns are: (1) object name; (2) jet viewing angle; 
  for the CLUMPY component: (3) redshift, (4) cloud optical depth, (5) 
  number of clouds along equatorial rays, (6) power-law index of the radial
  cloud distribution, (7) bolometric luminosity, (8) inner radius, and (9)
  mass in clouds (in solar masses). In all cases we assumed parameters
  $\sigma=45$ and $Y=10$.}

\medskip

\parbox[]{11cm}{$^{\star}$ objects presented in order of increasing jet viewing angle}

\end{table*}


The CLUMPY torus models are available for two different dust
compositions and have six parameters that can be adjusted: the optical
depth of the cloud, $\tau_{\rm V}$, the average number of clouds along
an equatorial line of sight, $N_0$, the power-law index of the radial
cloud distribution, $q$, the angular width of the Gaussian cloud
distribution, $\sigma$, the ratio between the outer and inner torus
radius, $Y$, and the viewing angle, $\theta$. The dust composition is
assumed to be a mix of silicates and graphite with silicates either as
in the standard interstellar medium \citep[][code ISM-D]{Draine03} or
with various materials added \citep[][code OHM]{OHM92}.

In general, the problem that one faces when choosing a CLUMPY model
that best-fits the data is that its parameters are highly degenerate,
with several permutations yielding very similar results
\citep{BayesClumpy09a, BayesClumpy09b, Nik09}. In the following, we
introduce an effective method to select a suitable model (by eye) and
find that the high degeneracy is due to the fact that, with the
exception of the dust composition, all parameters constrain only the
radial mass density profile (and so the radial temperature
profile). This then means that, although CLUMPY cannot uniquely
determine the AGN torus structure, it is a versatile tool with
possible application to other astrophysical objects.

As a first step, we find that our observations can constrain the dust
composition. As Fig. \ref{clumpydust} shows, ISM-D models approximate
well both the shape of the 10 $\mu$m silicate emission feature and the
spectral slope blueward of it, however, only if a redshift is applied
(top panels). On the other hand, OHM models predict a narrower
silicate feature than is observed and do not match well the blueward
spectral flux, independent of whether we apply a redshift or not
(bottom panels). We have then used in the following only ISM-D models.
In this respect, we note that our result differs from, but does not
contradict, that of \citet{Sir08}. These authors constrain the dust
composition of ultraluminous infrared galaxies based on the strengths
of both the 10 $\mu$m and 18 $\mu$m silicate features to be that of
the OHM models. However, a careful inspection of their Figs. 7-9 shows
that their method cannot constrain the dust composition in type-1 AGN,
such as our sources.

Next, we find that the most convenient and reliable way to constrain
the five geometrical CLUMPY parameters is to use the height of the 10
$\mu$m silicate emission peak relative to the 'dip' blueward of
it. Then, this flux ratio is larger, the smaller $N_0$, $q$, $\sigma$
and $\theta$, and the larger $Y$ are. The most pronounced differences
are produced by changing $N_0$, $q$, and $\sigma$
(Fig. \ref{clumpydens}), and the strongest degeneracy seems to be
present between $N_0$ and $\theta$ (i.e., we obtain a similar result,
if we either increase $N_0$ and decrease $\theta$ or vice
versa). Since all these parameters effectively determine the cloud
number density (see eq. (2) of \citet{Nen08b}), and the aforementioned
degeneracy serves to keep it constant, this then means that our flux
ratio is larger, the smaller $N_{\rm C}$. Once $N_{\rm C}$ is fixed,
the selection of $\tau_{\rm V}$ is straightforward, since models with
small values predict much stronger 10 $\mu$m silicate emission and
much less radiation at far-IR wavelengths than models with large
values \citep[see Fig. 5 of][]{Nen08b}.

\begin{figure*}
\centerline{
\includegraphics[scale=0.28]{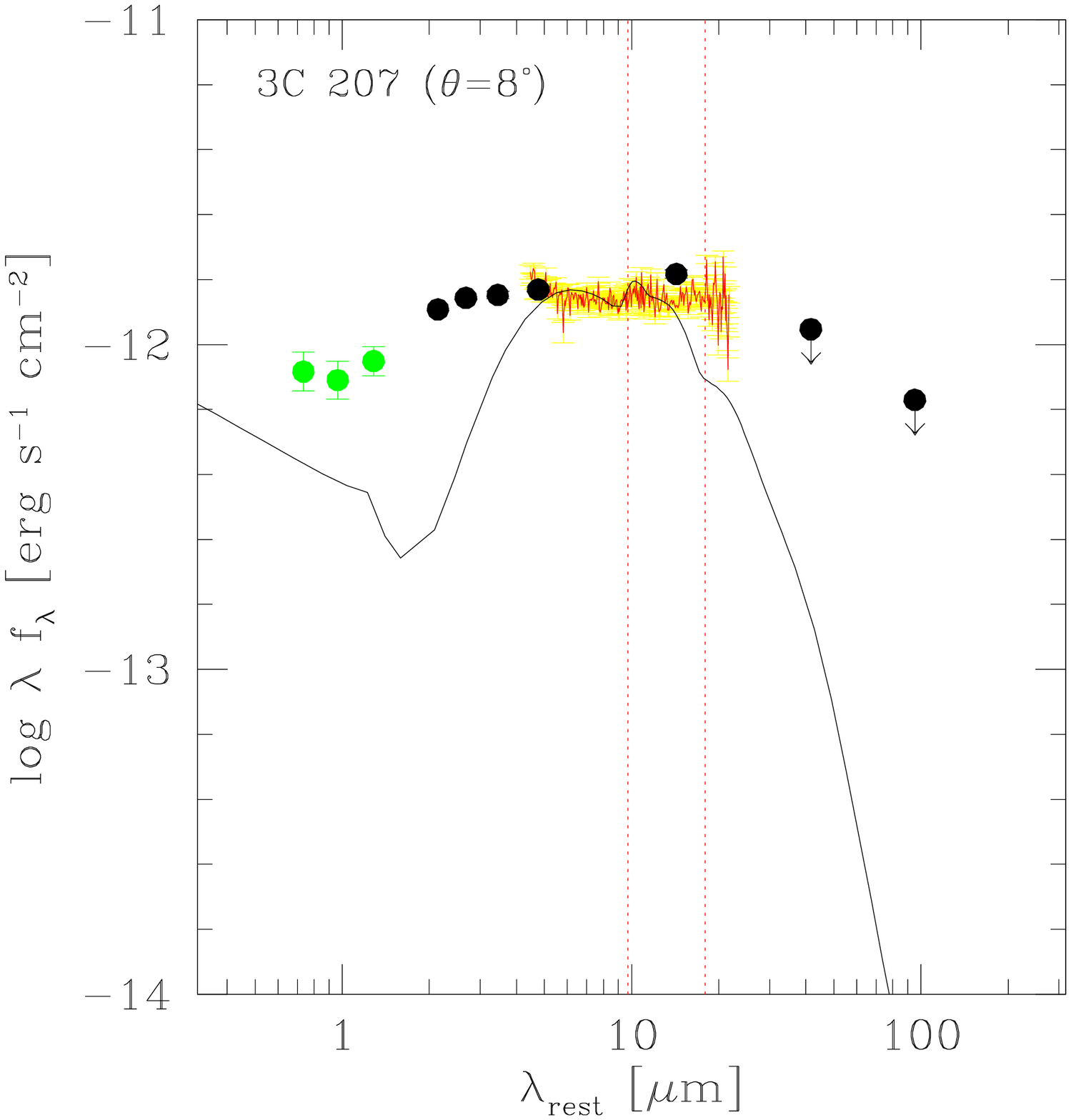} 
\includegraphics[scale=0.28]{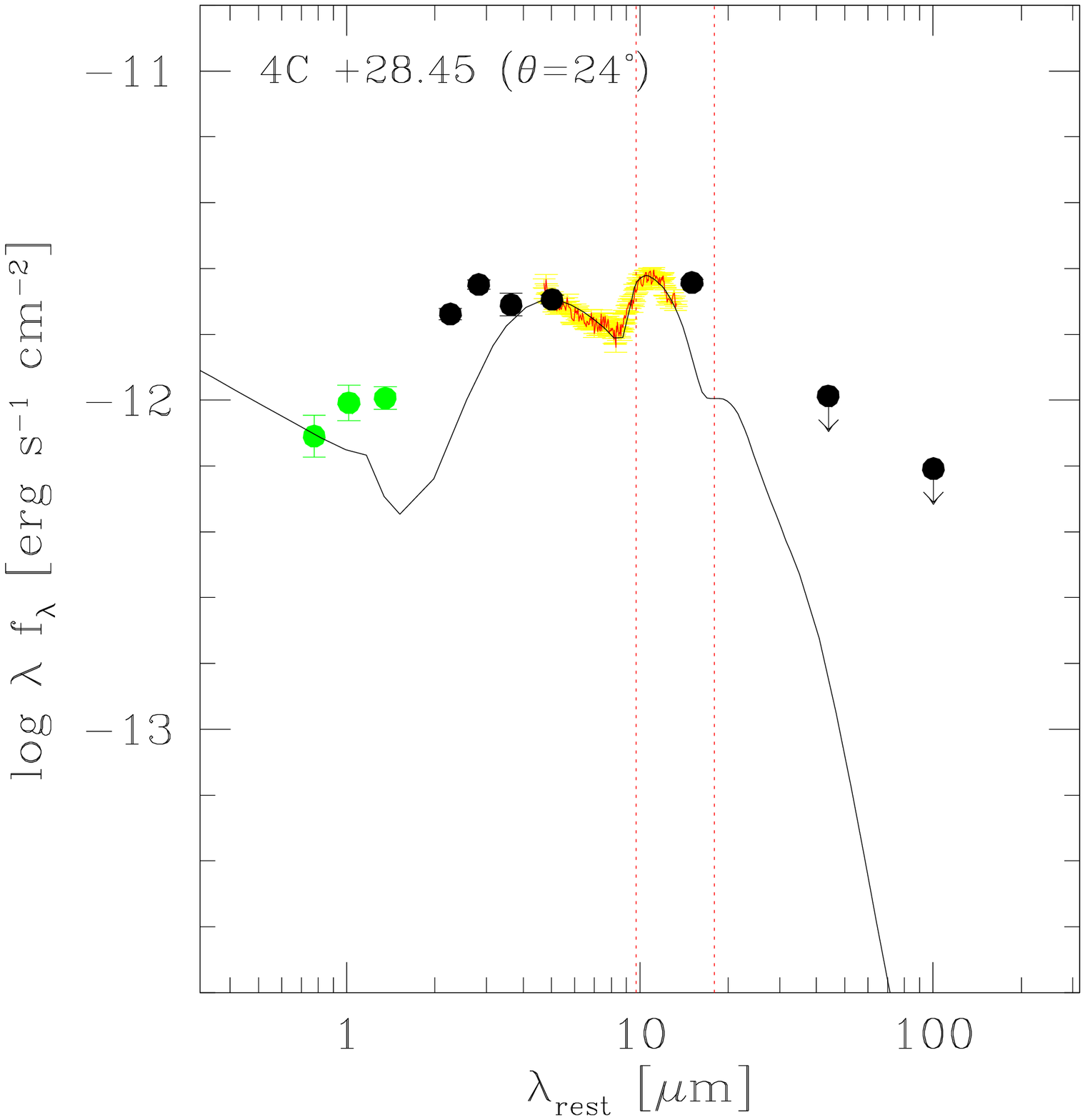}
\includegraphics[scale=0.28]{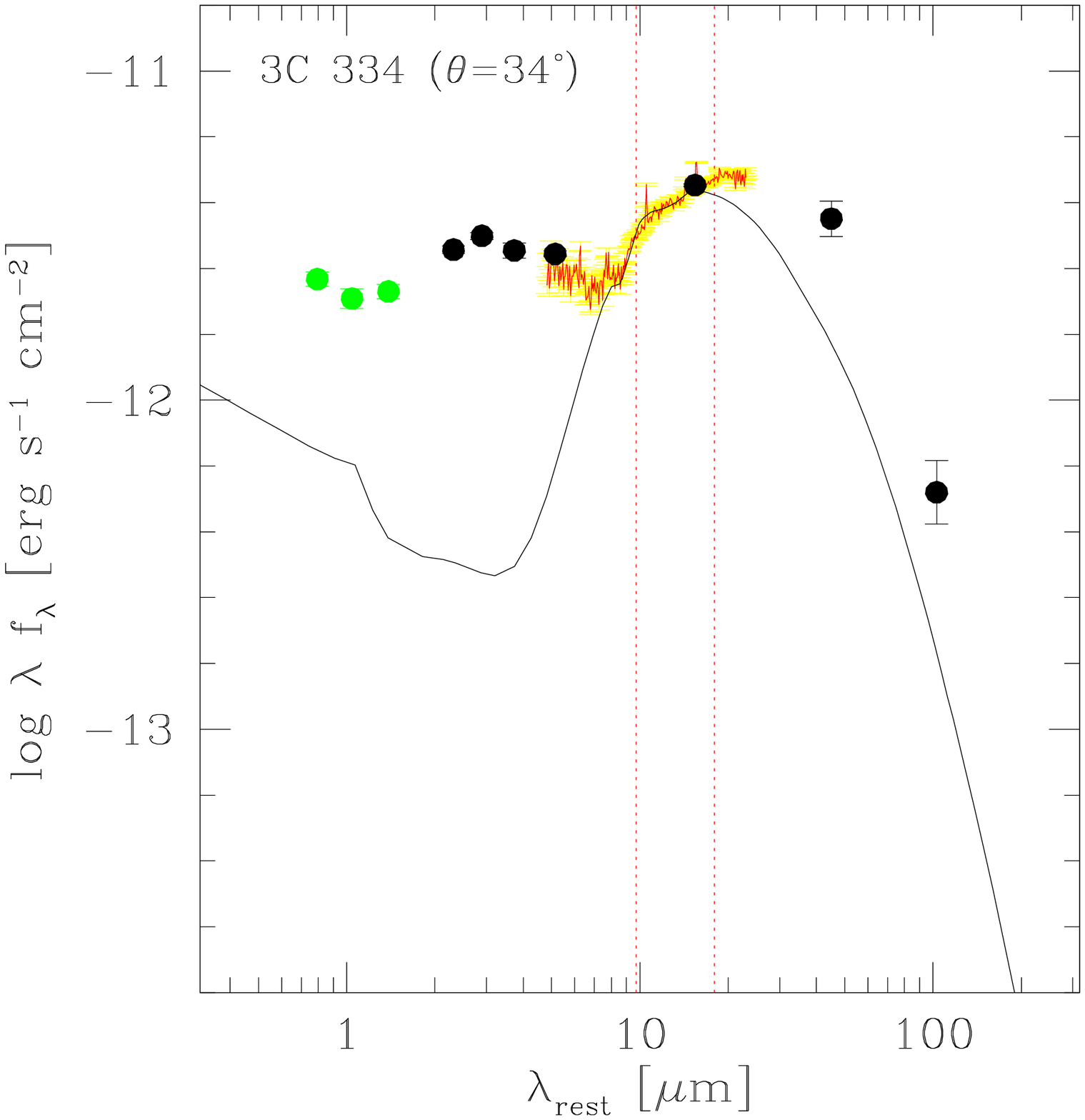}
}
\centerline{
\includegraphics[scale=0.28]{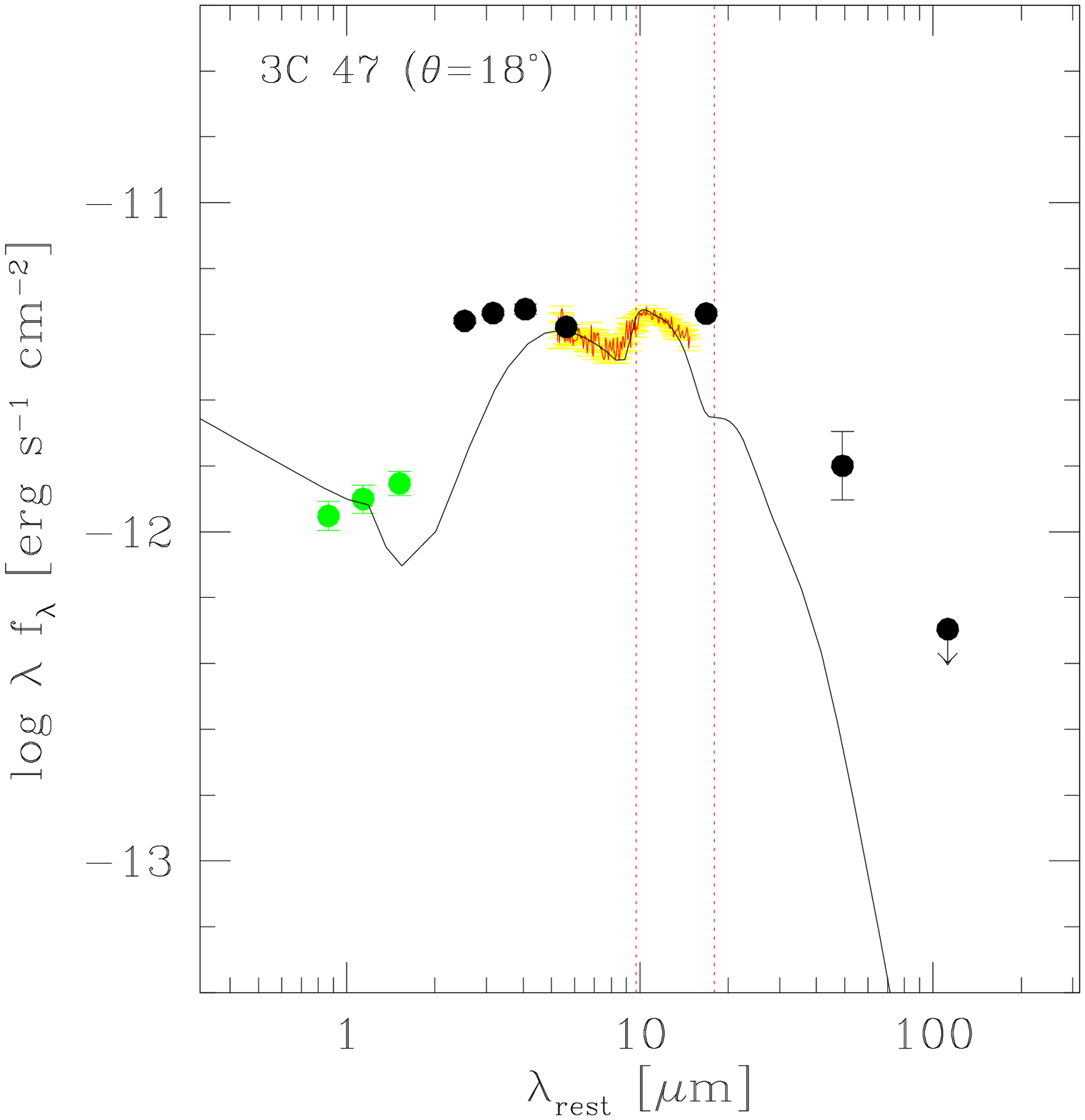}
\includegraphics[scale=0.28]{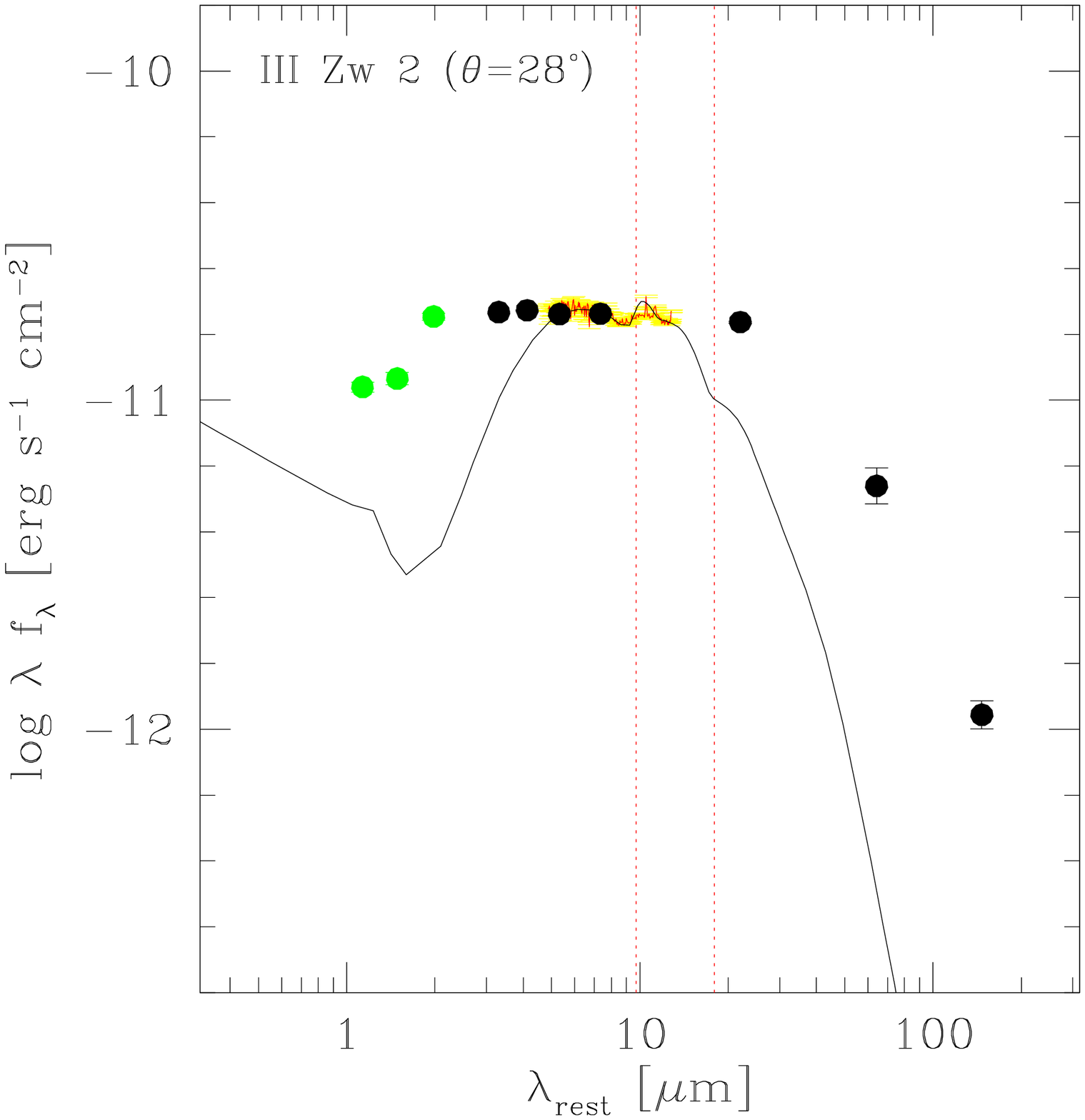}
\includegraphics[scale=0.28]{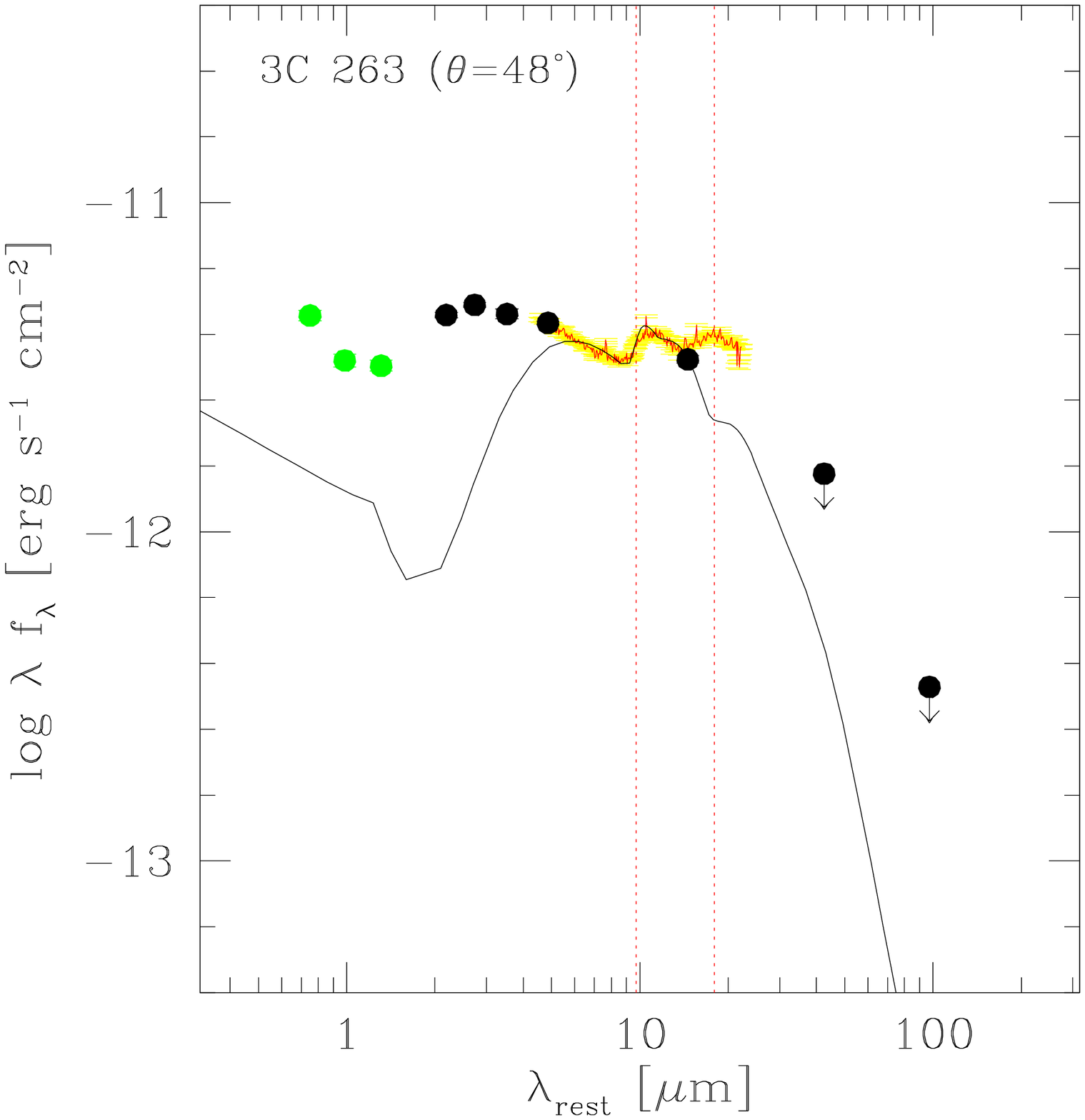}
}
\centerline{
\includegraphics[scale=0.28]{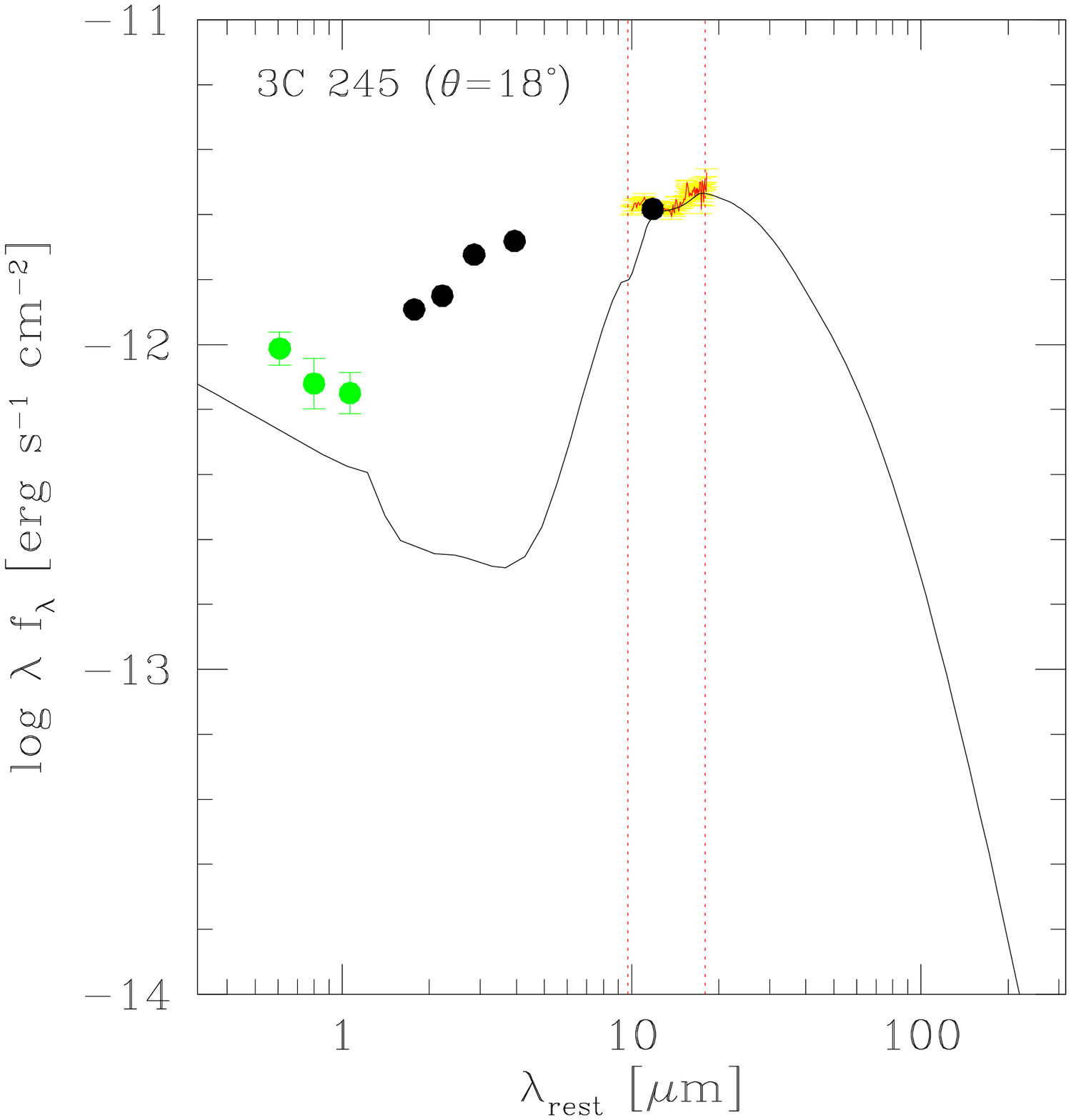}
\includegraphics[scale=0.28]{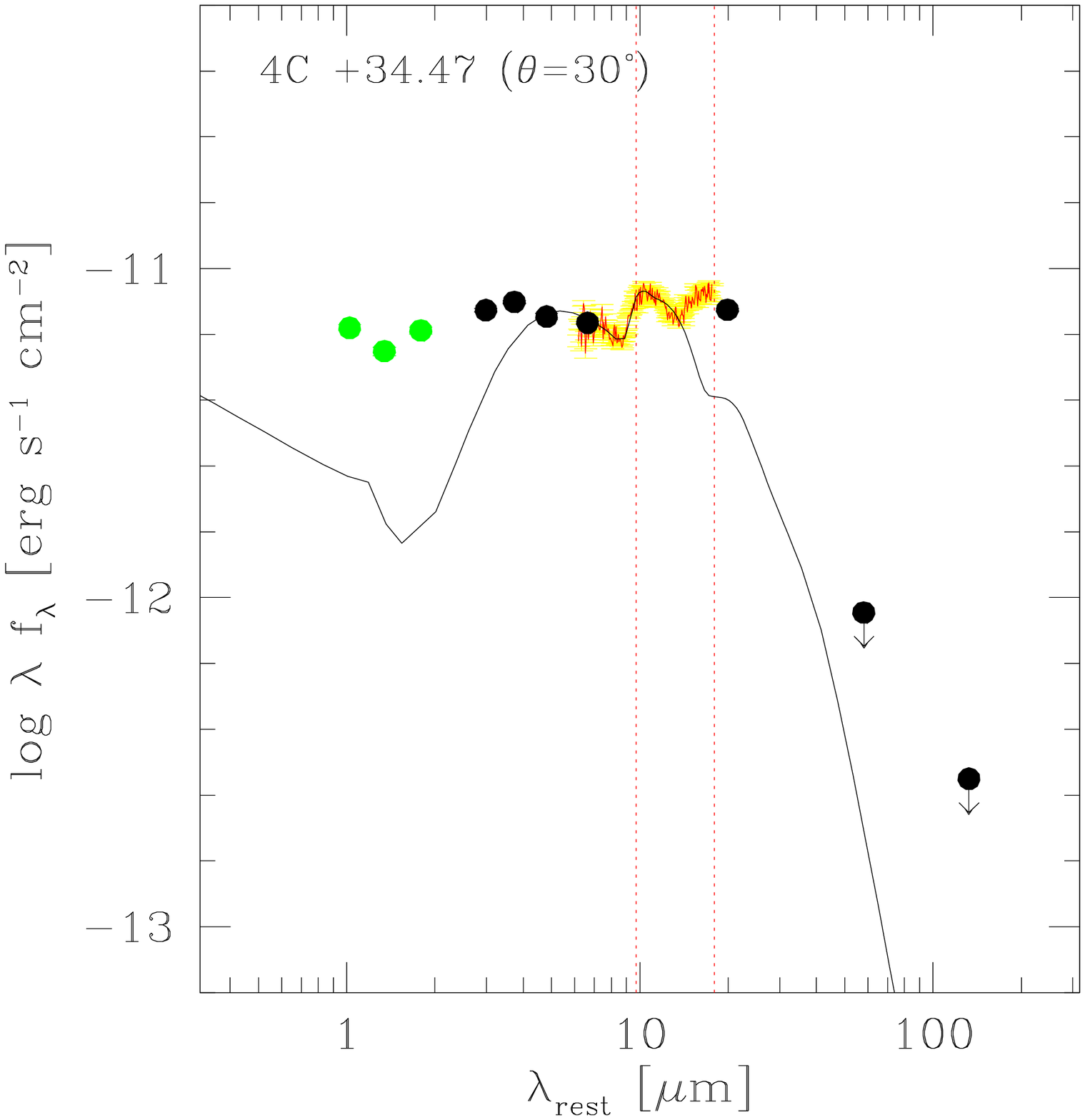}
\includegraphics[scale=0.28]{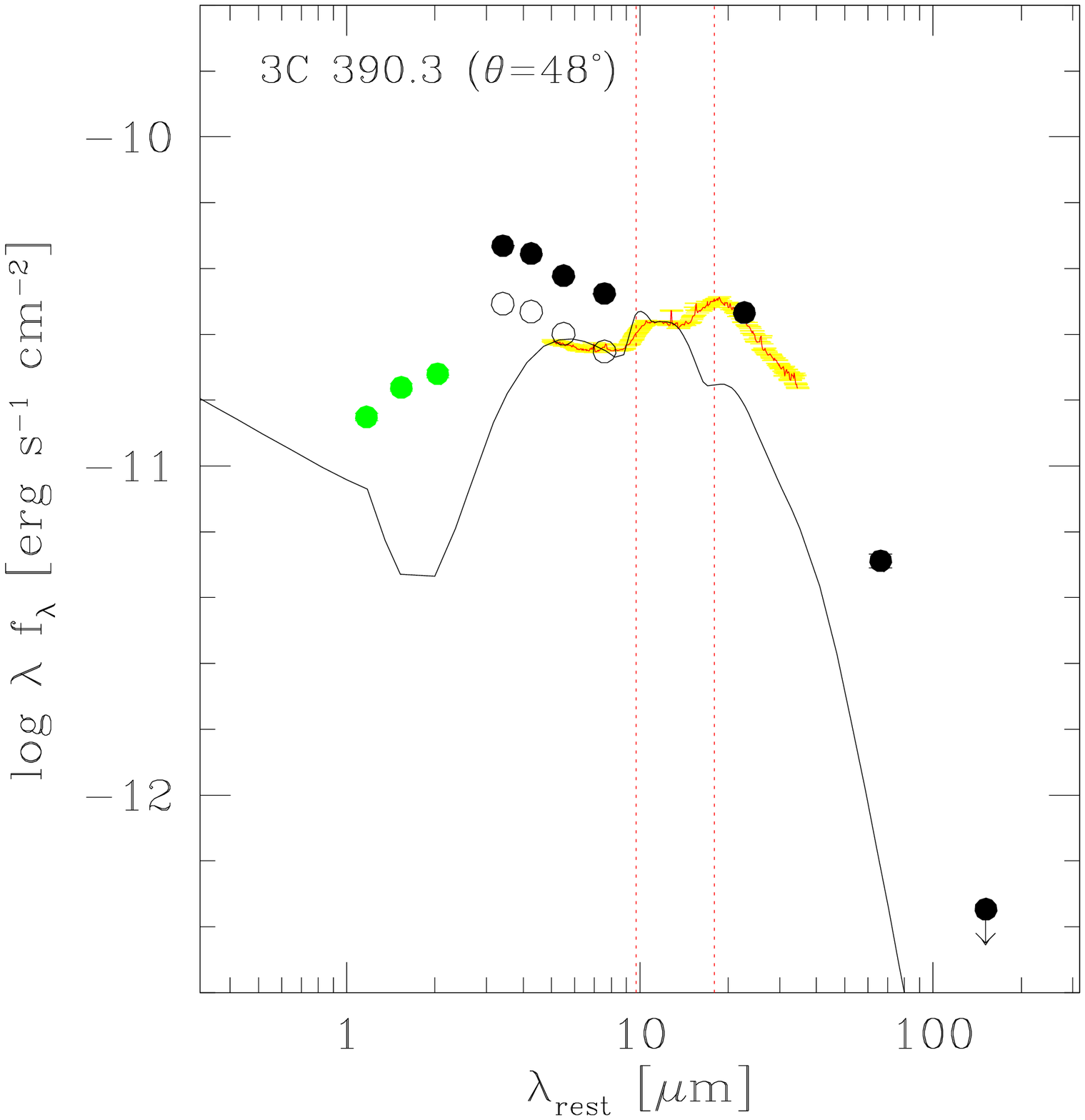}
}
\centerline{
\includegraphics[scale=0.28]{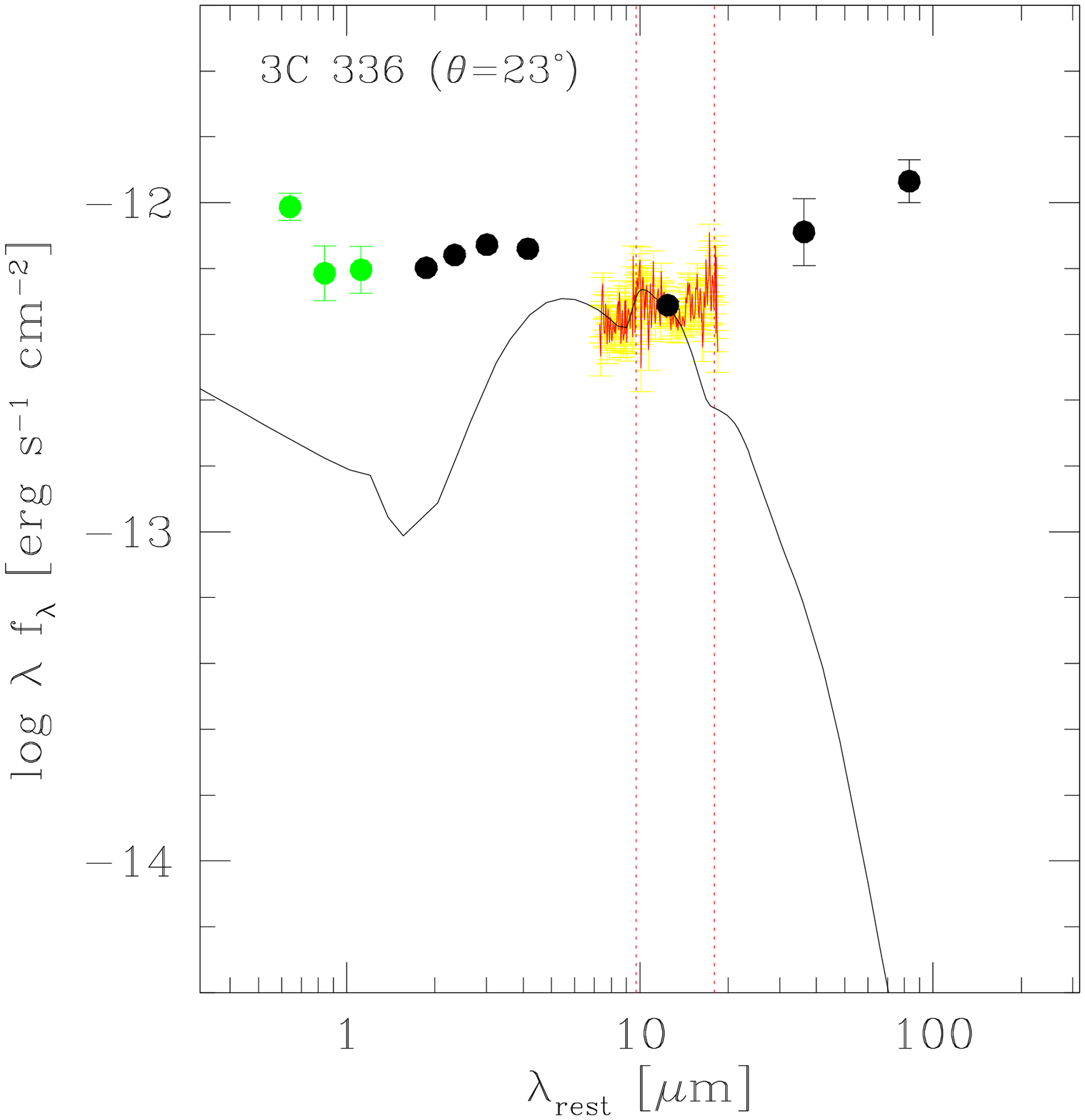} 
\includegraphics[scale=0.28]{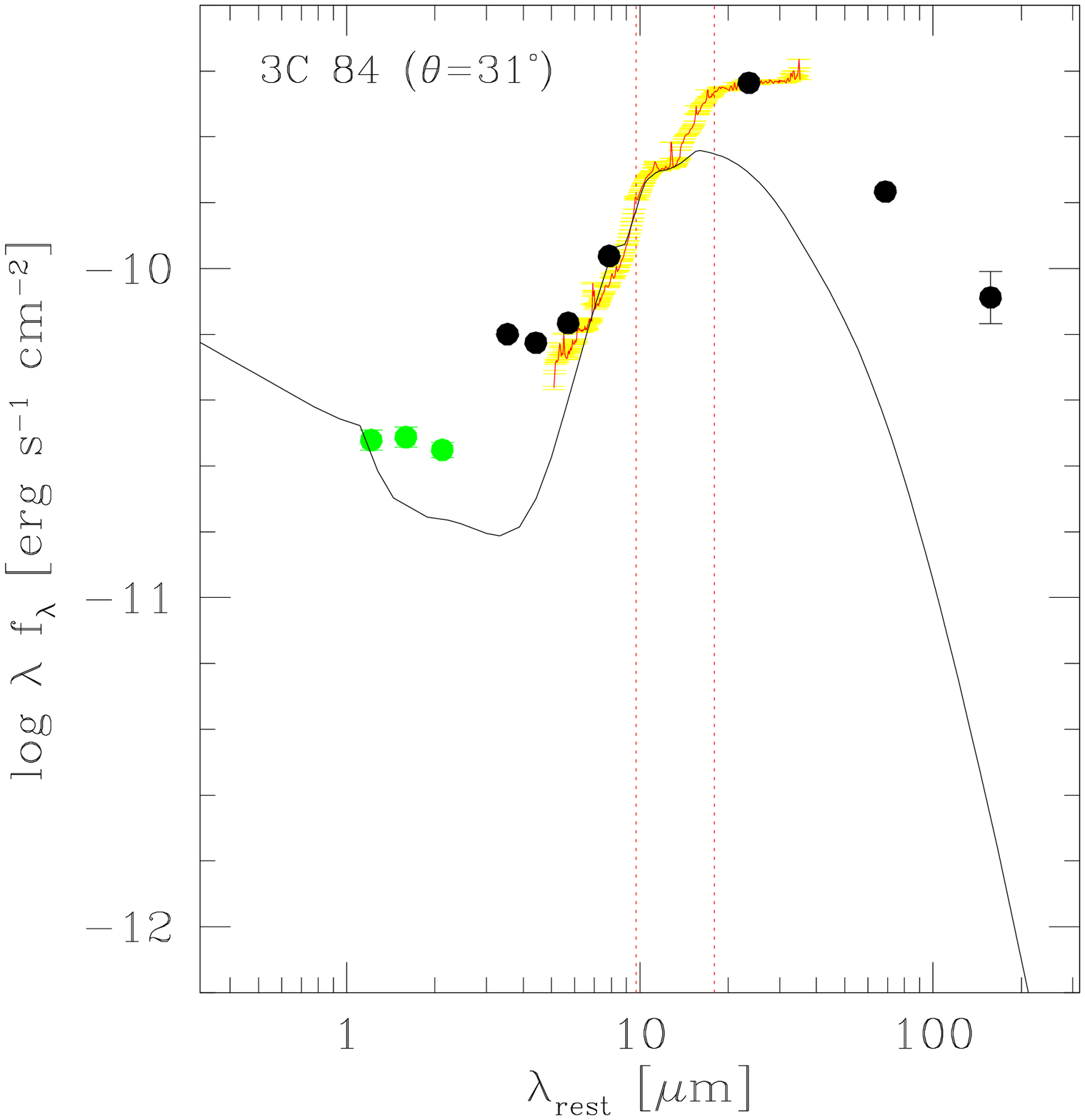}
\includegraphics[scale=0.28]{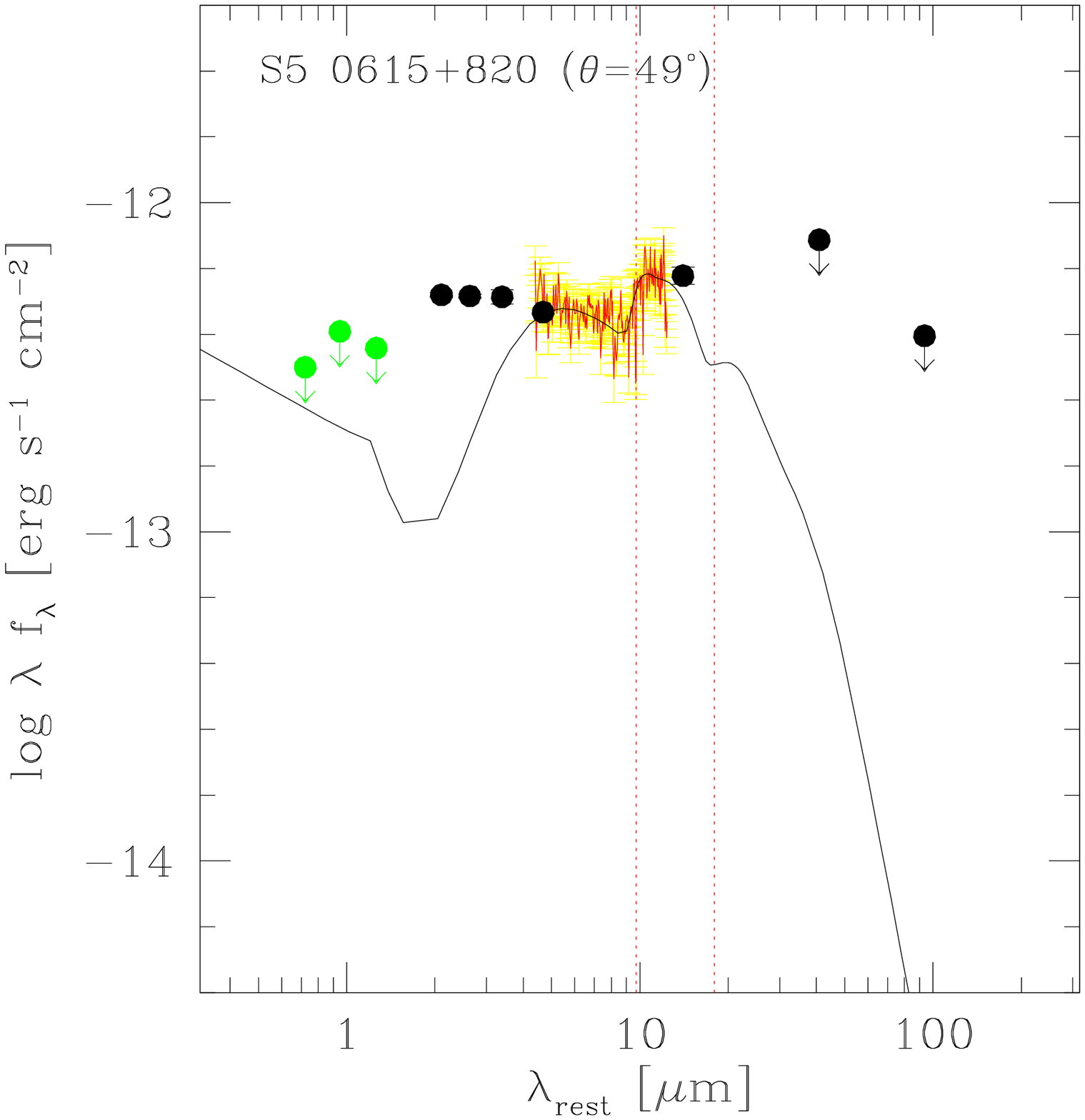}
}
\caption{\label{clumpysed} Infrared spectral energy distributions from
  Fig. \ref{sed} with {\it redshifted} CLUMPY torus models overlaid
  (solid black curves).}
\end{figure*}

\subsubsection{Revealing the strengths and deficits} \label{strengths}

Applying the CLUMPY models to the infrared SEDs of our sources, we
find in all cases that they can approximate well the mid-IR part, but
significantly underpredict the flux at both near-IR and far-IR
wavelengths. Furthermore, it is always necessary to apply a redshift
to the models. We show the models (including the predicted AGN
contribution) in Fig. \ref{clumpysed} (black solid curves) and list
the fitting set of parameters in Table \ref{clumpypar}.

The cold dust component unaccounted for by CLUMPY models is
ubiquitously observed in AGN and has been associated with dust from
either starbursts and/or the narrow emission line region (NELR)
\citep[e.g.,][]{Netzer07, Schweitzer08, Mor09}. The first alternative
is unlikely to hold for our sample for three reasons: (i) radio-loud
AGN reside predominantly in luminous ellipticals, which usually do not
have significant starbursts; (ii) we do not observe strong polycyclic
aromatic hydrocarbon (PAH) features in our IRS spectra, which are
usually associated with starbursts; and (iii) the observed peak
luminosities of the cold dust components are of the order of $L_{\rm
  cold} \sim 10^{10} - 10^{12}$ solar luminosities, implying that the
starburts would have to be unusually luminous. On the other hand,
models for the dust emission from the NELR usually predict higher
temperatures and, therefore, aim to account mainly for the wavelength
region covered by CLUMPY models \citep{Mor09}.

Besides the cold dust component, CLUMPY cannot account for the hottest
dust. The reason for this might be simply that its chemical
composition differs from that assumed by the models. For example, it
could be composed of pure graphite, which has a higher sublimation
temperature than silicates and emits a spectrum close to a pure
blackbody. Other authors have previously hinted at such an origin for
the hot dust component \citep[e.g.,][]{Mor09}.

\begin{figure}
\centerline{ 
\includegraphics[scale=0.4]{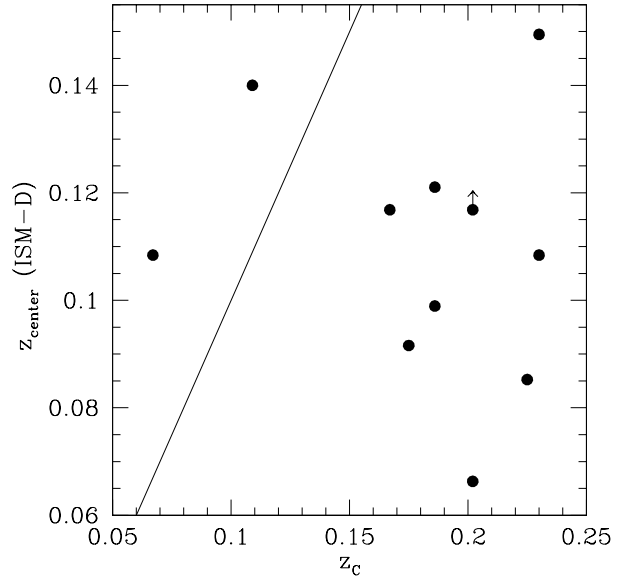}
}
\caption{\label{sired} The redshift of the center of the 10 $\mu$m
  silicate emission feature versus the redshift required to displace
  the CLUMPY models. The solid line marks the locus of
  equality. Arrows indicate limits.}
\end{figure}

\begin{figure}
\centerline{ 
\includegraphics[scale=0.4]{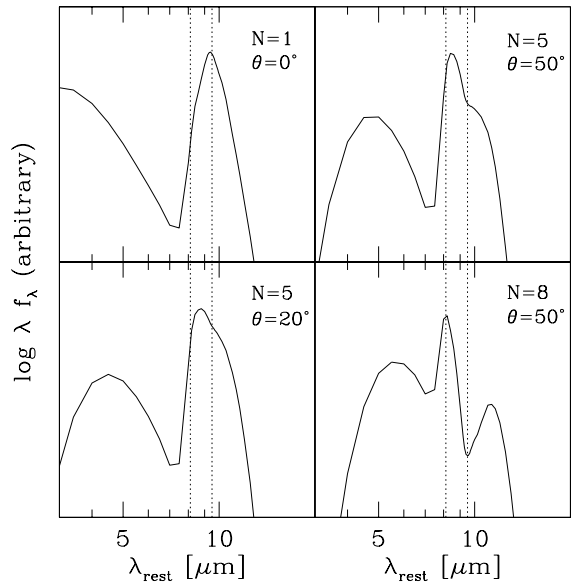}
}
\caption{\label{shift} The emission peak of the 10 $\mu$m silicate
  feature appears increasingly blueshifted as self-absorption becomes
  important. In the examples shown the feature center has shifted from
  9.5 $\mu$m (top left panel) to 8.1 $\mu$m (bottom right panel)
  (vertical dotted lines).}
\end{figure}

The redshifts required to displace the models are relatively large
($z_{\rm C} \sim 0.07 - 0.23$). Fig. \ref{sired} shows these redshifts
compared to those we derive from the measured center of the 10 $\mu$m
silicate emission feature (Table \ref{silicates}), with the latter
redshifts calculated assuming the peak rest-frame wavelength of the
ISM-D silicates (9.5 $\mu$m instead of 9.7 $\mu$m). With the exception
of the two sources that are modeled with a very high optical depth
(3C~84 and 3C~334, see below), which makes the determination of
$z_{\rm C}$ rather uncertain, the CLUMPY redshifts are always
larger. In order to understand this finding we recall that we
determine $z_{\rm C}$ by adjusting the CLUMPY model to the location of
the 'dip' blueward of the silicate feature and thus to the feature's
blue wing, whereas the measurement of the emission center depends
strongly on the width of the feature's top part. However, as CLUMPY
shows (Fig. \ref{shift}), the width of the top part is reduced as the
cloud number density and so the absorption increases, which leads to
an apparent blueshift of the emission peak and thus an underestimating
of the (true) redshift. This radiative transfer effect then also
offers an alternative explanation for the relation found in
Fig. \ref{silum}. Instead of a difference in grain size, it could be
that the larger the observed redshift of the center, the less absorbed
the silicate feature is, and so the larger its luminosity (for the
same AGN bolometric luminosity, as is the case for our sample; see
below).

Recently, \citet{Nik09} proposed that the observed redshifts of the
10~$\mu$m silicate emission features in AGN can be explained by an
interplay between the radiative transfer effects illustrated in
Fig. \ref{shift} and a rising continuum underneath the feature. In
particular, they showed that the appearance of the feature's top part
can change from less to more peaked depending on if the region between
the 10~$\mu$m and 18~$\mu$m silicate features was included in the
continuum fit or not. In this respect we note that we have included
the middle continuum part in our fit and have measured the center (not
the peak) of the feature, which is less sensitive to the continuum
placement. Furthermore, as Fig. \ref{shift} shows, the radiative
transfer effects mentioned by \citet{Nik09} {\it decrease} the
observed redshifts and so cannot explain them \citep[see
also][]{Hoenig10a}.

The large majority of our sources are best modelled assuming a cloud
optical depth of $\tau_{\rm V} = 10$, with only three objects (3C~245,
3C~84, and 3C~334) requiring a much larger value of $\tau_{\rm V} =
150$ in order to produce enough emission at large ($\lambda \ga
20~\mu$m) wavelengths. The number of clouds along equatorial rays and
the index of the radial cloud distribution differ the most among our
sources, with required values of $N_0 = 4, 5, 7$ and 20, and $q = 1,
2$ and 3, respectively. On the other hand, we find that it is not
necessary to vary the ratio between the outer and inner torus radius
and the width of the Gaussian distribution, with values of $Y=10$ and
$\sigma=45^\circ$, respectively, giving satisfactory results.

Several physical parameters can be derived from an approximation of
the infrared SED with CLUMPY models. The most important of these are
the bolometric luminosity of the source, the inner radius of the
CLUMPY component, and the observed mass in clouds (Table
\ref{clumpypar}). The bolometric luminosity can be readily obtained
from the scaling of the model to the data, taking into account that
the model is redshifted and thus the observed flux reduced by a factor
of $(1+z_{\rm C})^2$. The resulting values for our sources span only
$\sim 2$ orders of magnitude and we note that in particular the nine
objects that follow a relation in Fig. \ref{silum} have similar values
($\langle \log L_{\rm bol} \rangle = 45.9 \pm 0.1$ erg
s$^{-1}$). Using the bolometric luminosity and eq. (1) of
\citet{Nen08b} we have then calculated the inner radius of the CLUMPY
component, $R_{\rm in}$, assuming a dust sublimation temperature of
$T_{\rm sub} = 1500$ K. We obtain values in the range of $R_{\rm in}
\sim 0.2 - 2$ pc.

The calculation of the total mass in clouds follows eq. (6) of
\citet{Nen08a} and relies on the cloud number density, $N_{\rm C}$.
\citet{Nen08b} give an analytic expression for the mass in the case of
the sharp-edge geometry that differs only slightly from the proper
numerical integration in the case of a Gaussian toroidal geometry:

\begin{equation}
M_{\rm C} =  m_{\rm H} N_{\rm H} \int N_{\rm C}\, {\rm d}V = m_{\rm H} \frac{\tau_{\rm V}}{\sigma_{\rm V}} R_{\rm in}^2 N_0 I_{\rm q} 4 \pi \sin \sigma, 
\end{equation}

\noindent
where $m_{\rm H}$ is the proton mass, $N_{\rm H}$ is the hydrogen
column density of a cloud, $\sigma_{\rm V}$=4.89e$-$22 cm$^2$
\citep{Draine03} is the dust absorption cross-section in the $V$ band,
and $I_{\rm q} = (Y^2 - 1)/(2 \ln Y)$, $Y$, and $(2 \ln Y)/(1 -
Y^{-2})$ for the case of $q=1$, 2, and 3, respectively. Since the
degeneracy inherent to CLUMPY serves to keep the mass density
constant, the determination of $M_{\rm C}$ is expected to be robust,
i.e., one should get the same result with any fitting set of model
parameters.

\section{Summary and conclusions}

We have presented infrared observations obtained with all three
instruments on-board the {\it Spitzer Space Telescope} for 12
radio-loud AGN, for which actual viewing angles can be determined.
The results of our analysis of the infrared SED and the strength of
the 10 $\mu$m silicate feature can be summarized as follows:

\vspace*{0.2cm}

1. The infrared SED is best-fit with a combination of three or four
blackbodies. The resulting temperatures for the hot, warm, cool and
cold components are in the ranges $T_{\rm hot} \sim 1200 - 2000$ K,
$T_{\rm warm} \sim 300 - 800$ K, $T_{\rm cool} \sim 150 - 250$ K, and
$T_{\rm cold} \sim 60 - 150$ K, respectively.

2. We find trends between the emissions of the warm and cool dust
components and the jet viewing angle indicating that the more the
source is viewed face-on, the {\it larger} the amount of dust that
comes into our line of sight. No such trend is present for the hot
dust component. Both these results are contrary to the expectations of
smooth-density torus models.

3. Based on the overall continuum, we detect the 10 $\mu$m silicate
feature in {\it emission} in all our sources. The feature center is
always observed to be strongly redshifted.

4. We test the CLUMPY torus models of Nenkova and collaborators and
find that they approximate well the mid-infrared part of the SED, but
significantly underpredict the fluxes at both near-IR and far-IR
wavelengths. Furthermore, we find that the models can constrain the
dust composition (in our case to that of the standard ISM), that they
require relatively large redward displacements ($\sim 10\% - 20\%$ the
speed of light) to match the observations, and that they are
insensitive to the assumed geometry but give robust total mass
estimates.

\section*{Acknowledgments}

We thank Moshe Elitzur for valuable discussions. This work is based on
observations made with the Spitzer Space Telescope, which is operated
by the Jet Propulsion Laboratory, California Institute of Technology,
under a contract with the National Aeronautics Space Administration
(NASA). This research has made use of the NASA/IPAC Extragalactic
Database (NED), which is operated by the Jet Propulsion Laboratory,
California Institute of Technology, under contract with NASA.

\bibliography{references}

\bsp
\label{lastpage}

\end{document}